\documentclass[lineno]{jfm}

\usepackage{graphicx}
\usepackage{comment}
\usepackage{newtxtext}
\usepackage{newtxmath}
\usepackage{natbib}
\usepackage{hyperref}
\usepackage{bm}
\hypersetup{
    colorlinks = true,
    urlcolor   = blue,
    citecolor  = black,
}

\newcommand{\RomanNumeralCaps}[1]

\title{Statistical Nonlocality of Dynamically Coherent Structures}

\author{Andre N. Souza\aff{1}
  \corresp{\email{andrenogueirasouza@gmail.com}}, 
  Tyler Lutz\aff{2}, 
  \and 
  Glenn R. Flierl\aff{1}}

\affiliation{\aff{1}Massachusetts Institute of Technology, Cambridge, MA, United States}
\affiliation{\aff{2} Yale University, New Haven, CT, United States}

\newcommand{\avg}[1]{ \langle #1 \rangle }
\newcommand{\eref}[1]{ Equation \ref{#1} }
\begin{document}
\maketitle
\nolinenumbers 

\begin{abstract}
We introduce a class of stochastic advection problems amenable to analysis of turbulent transport. The statistics of the flow field are represented as a continuous time Markov process, a choice that captures the intuitive notion of turbulence as moving from one coherent structure to another. We obtain closed form expressions for the turbulent transport operator without invoking approximations. We recover the classical estimate of turbulent transport as a diffusivity tensor, the components of which are the integrated auto-correlation of the velocity field, in the limit that the operator becomes local in space and time.
\end{abstract}

\section{Introduction}
\label{sec:intro}

The study of passive scalar transport is of fundamental importance in characterizing turbulence. Given that predicting a chaotic dynamical trajectory over long times is infeasible \citep{Lorenz1963} one must switch to a statistical perspective to make headway on transport properties. From the analysis of anomalous dispersion by \cite{Taylor1920}, operator notions of mixing from \cite{knobloch_1977}, computations of ``effective diffusivity" by \cite{Majda1991}, simplified models of turbulence by \cite{Pope2011}, rigorous notions of mixing in terms of Sobolev norms by \cite{Thiffeault2012}, or upper bounds on transport as in \cite{hassanzadeh2014}, different approaches elucidate fundamental properties of turbulence.

In addition to furthering our understanding of turbulence, there are practical applications for turbulence closures. In particular, Earth Systems Models require closure relations for the transport of unresolved motions \cite{Scheider2017}; however, the closure relations are marred by structural and parametric uncertainty, requiring ad-hoc tuning to compensate for biases. There are structural biases associated with scaling laws and closure assumptions between turbulent fluxes and gradients. Modern studies are bridging the gap by incorporating more complex physics and novel scaling laws, \cite{Teixeira2018, Gallet2020}, but the chosen functional forms to represent fluxes remain unknown. 

The multi-scale nature of turbulent flows, the presence of coherent structures,  as well as the interconnection of reacting chemical species and prognostic fields, instead suggests that fluxes are perhaps better modeled using nonlinear nonlocal (in space, time, and state) operators. Data-driven methods relying on flexible interpolants can significantly reduce structural bias, but often at the expense of interpretability, generalizability, or efficiency. Thus, understanding scaling laws and functional forms of turbulence closures is still necessary to physically constrain data-driven methods and decrease their computational expense. A promising avenue for significant progress, lying at the intersection of theory and practice, is the calculation of closure relations for passive scalars.

The present work aims to characterize the transport of passive scalars by flow fields with known statistics. Our notion of transport is the ensemble mean flux, which we express as an operator that acts on the ensemble mean tracers in terms of the statistics of the flow field. We make arguments akin to those in \cite{Kraichnan1968} to motivate the operator approach, but our method of calculation is fundamentally different. Given that the goal is to construct an operator, rather than estimating a diffusivity tensor acting on the ensemble mean gradients or deriving upper bounds, we take a field-theoretic perspective \citep{Hopf1952}. Doing so allows us to derive a coupled set of partial differential equations representing conditional mean tracers where the conditional averages are with respect to different flow states. The turbulent transport operator is then a Schur complement of the resulting linear system with respect to statistical ``perturbation" variables. If the flow statistics are given by a continuous-time Markov process with a small finite state space, the Schur complement becomes tractable to compute analytically. Obtaining a closed form functional for turbulent fluxes allows for a direct statistical simulation approach similar to those of \cite{Karniadakis2013, Marston2016, Farrell2019}.

The paper is organized as follows. In section \ref{sec:theory} we formulate the closure problem and recast it as one of solving coupled partial differential equations. In section \ref{sec:examples_section} we show how to explicitly solve the closure problem for a few flows with simple temporal structure but arbitrarily complex spatial structure. In section \ref{sec:general_theory} we outline the general theory. Appendices supplement the body of the manuscript. Appendix \ref{FeynmanStyle} provides a direct field-theoretic derivation of arguments in section \ref{sec:theory} and Appendix \ref{heuristic_overview} provides a heuristic overview of obtaining continuous-time Markov processes and their statistics from deterministic or stochastic dynamical systems.

\section{Problem Formulation}
\label{sec:theory}
We consider the advection and diffusion of an ensemble of passive scalars $\theta_{\omega}$ 
\begin{align}
    \partial_t \theta_{\omega} +  \nabla \cdot \left( \bm{u}_{\omega}  \theta_{\omega} - \kappa \nabla \theta_\omega \right) &=  s(\bm{x} )
\end{align}
by a stochastic flow field $\bm{u}_{\omega}(\bm{x}, t)$ where $\omega$ labels the ensemble member. Here $s$ is a deterministic mean zero source term and $\kappa$ is a diffusivity constant\footnote{For laboratory flows $\kappa$ would be the molecular diffusivity; for larger-scale problems, we rely on the fact that away from boundaries, the ensemble mean advective flux (but not necessarily other statistics) may still be much larger than the diffusive flux and thus the exact value of $\kappa$ will not matter.}.
 Our target is to obtain a meaningful equation for the ensemble mean,
\begin{align}
\label{mean_equations}
    \partial_t \avg{\theta} + \nabla \cdot\left(  \avg{ \bm{u}  \theta } - \kappa \nabla \avg{ \theta } \right) &=  s(\bm{x})
\end{align}
which requires a computationally amenable expression for the mean advective flux, $\avg{ \bm{u} \theta } $, in terms of the statistics of flow field, $\bm{u}(\bm{x}, t)$, and the ensemble average of the tracer, $\avg{\theta}$. Thus the closure problem is to find an operator $\mathcal{O}$ that relates the ensemble mean, $\langle \theta \rangle$, to the ensemble mean advective-flux, $\langle \bm{u} \theta \rangle $, i.e., 
\begin{align}
    \mathcal{O}[\langle \theta \rangle] = \langle \bm{u} \theta \rangle .
\end{align}
We show how to define (and solve for) the operator $\mathcal{O}$. The operator will be linear with respect to its argument and depend on the statistics of the flow field. 

We assume all tracer ensemble members to have the same initial condition and thus the ensemble average here is with respect to different flow realizations. The only source of randomness comes from different flow realizations. Throughout the manuscript we assume homogeneous Neumann boundary conditions for the tracer and zero wall-normal flow for the velocity field when boundaries are present. These restrictions, combined with the assumption that the source term is mean zero, imply that the tracer average is conserved.

For the statistics of the flow field, we consider a continuous time Markov process with $N$ states corresponding to steady flow fields $\bm{u}_n(\bm{x})$ where $n$ is the associated state index. We start with finitely many states for simplicity rather than necessity. Physically, we think of these states as representing coherent structures in a turbulent flow\footnote{This is viewed as a finite volume discretization in function space where the states are the ``cell averages" of a control volume in function space.}. A turbulent flow, by its very nature, chaotic and unpredictable over long time horizons, is modelled as being semi-unpredictable through our choice. Over short horizons, the probability of remaining in a given state is large. On the medium term, the flow is limited to moving to a subset of likely places in phase space. Over long time horizons, the most one can say about the flow is related to the likelihood of being found in the appropriate subset of phase space associated with the statistically steady state. 

Thus we proceed by characterizing the probability, $\mathbb{P}$, of transitioning from state $n$ to state $m$ by a transition matrix $\mathscr{P}(\tau)$,
\begin{align}
\mathbb{P}\{ \bm{u}(\bm{x}, t+\tau) = \bm{u}_m(\bm{x}) | \bm{u}(\bm{x}, t) = \bm{u}_n(\bm{x}) \} &= [\mathscr{P}(\tau)]_{mn}.
\end{align}
The transition probability is defined through its relation to the generator $\mathcal{Q}$,
\begin{align}
\mathscr{P}(\tau) \equiv \exp( \mathcal{Q} \tau ) 
\end{align}
where $\exp( \mathcal{Q} \tau )$ is a matrix exponential. Each entry of $\mathscr{P}(\tau)$ must be positive. Furthermore, the column sum of $\mathscr{P}(\tau)$ for each $\tau$, sum to one since the total probability must sum to one. Similarly $\mathcal{Q}$'s off-diagonal terms must be positive\footnote{Indeed, to first order $\exp(\mathcal{Q} dt) =  \mathbb{I} + \mathcal{Q} dt$. The positivity requirement of the transition probability $\mathscr{P}(dt)=\exp(\mathcal{Q} dt)$ necessitates the positivity of $\mathcal{Q}$'s off-diagonal terms as well as the negativity of the diagonal terms.} and the column sum of $\mathcal{Q}$ must be zero. 

We denote the probability of being found at state $m$ at time $t$ by $\mathcal{P}_m(t)$,
\begin{align}
    \mathcal{P}_m(t) = \mathbb{P}\{\bm{u}(\bm{x}, t) = \bm{u}_m(\bm{x}) \} .
\end{align}
The evolution equation for $\mathcal{P}_m(t)$ is the master equation,
\begin{align}
\label{master_equation_u}
\frac{d}{dt} \mathcal{P}_m &= \sum_n \mathcal{Q}_{mn} \mathcal{P}_n .
\end{align}
We assume that Equation \ref{master_equation_u} has a unique steady state and denote the components of the steady state by $P_m$. 

We have used several "P"s at this stage and their relation are:
\begin{enumerate}
    \item $\mathbb{P}$ denotes a probability. 
    \item $\mathscr{P}(\tau)$ denotes the transition probability matrix for a time $\tau$ in the future.
    \item $\mathcal{P}_m(t)$ denotes the probability of being in state $m$ at time $t$. The algebraic relation \[ \sum_m P_m(t+\tau) \bm{\hat{e}}_m = \mathscr{P}(\tau) \sum_n P_n(t) \bm{\hat{e}}_n \] holds. 
    \item $P_m$ is the statistically steady probability of being found in state $m$. In the limit \[ \lim_{t \rightarrow \infty} \mathcal{P}_m(t) = P_m . \] 
\end{enumerate}

We exploit the given information about the flow field to infer the mean statistics of the passive tracer $\theta_\omega$. We do so by conditionally averaging the tracer field $\theta_\omega$ with respect to a given flow state $\bm{u}_n$. More precisely, given the stochastic partial differential equation,
\begin{align}
\label{stochastic_system_u}
\mathbb{P}\{ \bm{u}_\omega( \bm{x}, t+\tau) = \bm{u}_m(\bm{x}) | \bm{u}_\omega(\bm{x}, t) = \bm{u}_n(\bm{x}) \} &= [\exp( \mathcal{Q} \tau )]_{mn},
\\
\label{stochastic_system_theta}
    \partial_t \theta_\omega +  \nabla \cdot \left( \bm{u}_\omega \theta_\omega - \kappa \nabla \theta_\omega \right) &=  s( \bm{x} ) ,
\end{align}
we shall obtain equations for probability weighted conditional means of $\theta_\omega$ defined by
\begin{align}
\Theta_m(\bm{x} , t) &\equiv  \langle \theta_\omega \rangle_{\bm{u}(\bm{x}, t) = \bm{u}_m(\bm{x})} \mathcal{P}_m(t) .
\end{align}
We will show that the evolution equation for $\Theta_m$ is 
\begin{align}
\label{conditional_mean_equation_P_preview}
\frac{d}{dt} \mathcal{P}_m &= \sum_n \mathcal{Q}_{mn} \mathcal{P}_n , \\
\label{conditional_mean_equation_theta_preview}
    \partial_t \Theta_m + \nabla \cdot \left( \bm{u}_m \Theta_m - \kappa \nabla \Theta_m \right) &= s(\bm{x}) \mathcal{P}_m  + \sum_n \mathcal{Q}_{mn} \Theta_n .
\end{align}
The explicit dependence on the generator in Equation \ref{conditional_mean_equation_theta_preview}, as we shall see, yields considerable information. We recover the equation for the tracer ensemble mean, Equation \ref{mean_equations}, by summing Equation \ref{conditional_mean_equation_theta_preview} over the index $m$, using $\langle \theta \rangle = \sum_m \Theta_m$, $\sum_m \mathcal{Q}_{mn} = \bm{0}$, and $\sum_m \mathcal{P}_{m} = 1$, 
\begin{align}
    \partial_t \sum_m \Theta_m + \nabla \cdot \left( \sum_m \bm{u}_m \Theta_m - \kappa \nabla \sum_m \Theta_m \right) &= s(\bm{x})  \\
    \nonumber
    &\Leftrightarrow \\
    \partial_t \langle \theta \rangle  + \nabla \cdot \left( \langle \bm{u} \theta \rangle - \kappa \nabla \langle \theta \rangle \right) &= s(\bm{x})  .
\end{align}
We comment that the presence of the generator when taking conditional averages is similar to the entrainment hypothesis in the atmospheric literature. See, for example, \cite{Teixeira2018} for its use in motivating a turbulence closure; however, here we derive the result from the direct statistical representation as opposed to hypothesize its presence from a dynamical argument.

Most of the terms in Equation \ref{conditional_mean_equation_theta_preview} are obtained by applying a conditional average to Equation \ref{stochastic_system_theta}, commuting with spatial derivatives when necessary, and then multiplying through by $\mathcal{P}_m$; however, the primary difficulty lies in proper treatment of the conditional average of the temporal derivative. We circumvent the problem in a roundabout manner: The strategy is to discretize the advection-diffusion equation, write down the resulting master equation, compute moments of the probability distribution, and then take limits to restore the continuum nature of the advection-diffusion equation. For an alternative derivation where we forego discretization see Appendix \ref{FeynmanStyle} and for a brief overview of the connection between the discrete, continuous, and mixed master equation see Appendix \ref{heuristic_overview} or, in a simpler context, \cite{Doering_1989}.

A generic discretization (in any number of dimensions) of Equation \ref{stochastic_system_theta} is of the form 
\begin{align}
\label{discretized_theta}
\frac{d}{dt} \theta^i + \sum_{jkc} A_{ijk}^c u^{k,c}_\omega \theta^j  - \sum_j D_{ij} \theta^j &=  s^i  
\end{align}
for some tensor $A_{ijk}^c$, representing advection, and matrix $D_{ij}$, representing diffusion. Here each $i,j$ and $k$ corresponds to a spatial location, and the index $c$ corresponds to a component of the velocity field $\bm{u}$. The variable $\theta^i$ is the value of the tracer at grid location $i$ and $u^{k,c}$ is the value of the $c$'th velocity component and grid location $k$. The master equation for the joint probability density for each component $\theta^i$ and Markov state $m$, $\rho_m( \bm{\theta} )$, where $\bm{\theta} = (\theta^1, \theta^2, ....)$ and the $m$-index denotes a particular Markov state, is a combination of the Liouville equation for \ref{discretized_theta} and the transition rate equation for \ref{stochastic_system_u},
\begin{align}
\label{discrete_master_equation}
    \partial_t \rho_m &= \sum_{i} \frac{\partial}{\partial \theta^i} \left[ \left( \sum_{jkc} A_{ijk}^c u^{k,c}_m \theta^j  - \sum_j D_{ij} \theta^j -  s^i  \right ) \rho_m \right] + \sum_n \mathcal{Q}_{mn} \rho_n .
\end{align}
Define the following moments,
\begin{align}
\label{prob_1}
\mathcal{P}_m &= \int d \bm{\theta} \rho_m \text{ and }
\Theta_m^j = \int d \bm{\theta} \theta^j \rho_m  .
\end{align}
We obtain an equation for $\mathcal{P}_m$ by integrating \ref{discrete_master_equation} by $d \bm{\theta}$ to yield 
\begin{align}
\frac{d}{dt} \mathcal{P}_m &= \sum_n \mathcal{Q}_{mn} \mathcal{P}_n
\end{align}
as expected from Equation \ref{master_equation_u}. The equation for $\Theta_m^\ell$ is obtained by multiplying \ref{discrete_master_equation} by $\theta^\ell$ and then integrating with respect to $d \bm{\theta}$,
\begin{align}
\label{discrete_master_equation_conditional_theta}
    \frac{d}{dt} \Theta_m^\ell &=   -\sum_{jkc} A_{\ell jk}^c u^{k,c}_m \Theta^j_m + \sum_j D_{\ell j} \Theta^j_m  + s^\ell \mathcal{P}_m  + \sum_n \mathcal{Q}_{mn} \Theta_n^\ell ,
\end{align}
where we integrated by parts on the $\int d \bm{\theta} \theta^\ell \partial_{\theta^i} \bullet $ term. 
Upon taking limits of \ref{discrete_master_equation_conditional_theta} we have the following equations
\begin{align}
\label{conditional_mean_equation_P}
\frac{d}{dt} \mathcal{P}_m &= \sum_n \mathcal{Q}_{mn} \mathcal{P}_n \\
\label{conditional_mean_equation_theta}
    \partial_t \Theta_m + \nabla \cdot \left( \bm{u}_m \Theta_m - \kappa \nabla \Theta_m \right) &= s(\bm{x}) \mathcal{P}_m  + \sum_n \mathcal{Q}_{mn} \Theta_n.
\end{align}
We compare Equation \ref{conditional_mean_equation_theta} to the direct application of the conditional average to Equation \ref{stochastic_system_theta} followed by multiplication with $\mathcal{P}_m$ to infer,
\begin{align}
\label{time_relation}
\langle \partial_t \theta_\omega \rangle_{\bm{u}(\bm{x},t) = \bm{u}_m(\bm{x}) } \mathcal{P}_{m} &= \partial_t \Theta_m - \sum_{n} \mathcal{Q}_{mn} \Theta_n .
\end{align}

In summary, for an m-dimensional advection diffusion equation and $N$ Markov states, Equations \ref{conditional_mean_equation_P_preview}-\ref{conditional_mean_equation_theta_preview} are a set of $N$-coupled m-dimensional advection diffusion equations with $N$ different steady velocities. When the statistics of the flow field are described by $c$ continuous variables, the resulting equation set becomes an $m+c$ dimensional system. Stated differently, if the statistics of $\bm{u}_\omega$ are characterized by a transitions between a continuum of states associated with a linear operator $\mathcal{F}_{\bm{\omega}}$ with variables $\bm{\omega} \in \mathbb{R}^c$, then 
\begin{align}
\label{continuous_conditional_mean_equation_P}
    \partial_t \mathcal{P} &= \mathcal{F}_{\bm{\omega}}[\mathcal{P}] \\
\label{continuous_conditional_mean_equation_theta}
    \partial_t \Theta + \nabla \cdot \left( \bm{u} \Theta - \kappa \nabla \Theta \right) &= s(\bm{x}) \mathcal{P}  + \mathcal{F}_{\bm{\omega}}[ \Theta],
\end{align}
where $\mathcal{P} = \mathcal{P}(\bm{\omega}, t)$,  $\Theta = \Theta(\bm{x}, \bm{\omega}, t)$, and $\bm{u} = \bm{u}(\bm{x}, \bm{\omega})$. Equations \ref{conditional_mean_equation_P}-\ref{conditional_mean_equation_theta} are thought of as finite volume discretizations of flow statistics in Equations \ref{continuous_conditional_mean_equation_P}-\ref{continuous_conditional_mean_equation_theta}.

Our primary concern in this work is to use Equations \ref{conditional_mean_equation_P_preview}-\ref{conditional_mean_equation_theta_preview} to calculate meaningful expressions for $\langle \bm{u} \theta \rangle$; however, we shall first take a broader view to understand the general structure of the turbulent fluxes. The following argument is attributed to \cite{Weinstock1969}, but we use our own notation and make additional simplifications.

Applying the Reynolds decomposition 
\begin{align}
\theta_\omega = \langle \theta \rangle + \theta'_\omega \text{ and  }   \bm{u}_\omega = \langle \bm{u} \rangle + \bm{u}'_\omega
\end{align}
yields 
\begin{align}
\partial_t \langle \theta \rangle + \nabla \cdot \left( \langle \bm{u} \rangle \langle \theta \rangle + \langle \bm{u}' \theta' \rangle - \kappa \nabla \langle \theta \rangle \right) &= s \\
 \partial_t \theta'_\omega + \nabla \cdot \left( \langle \bm{u} \rangle \langle \theta \rangle - \langle \bm{u}' \theta' \rangle  + \bm{u}_\omega \theta_\omega - \kappa \nabla \theta'_\omega \right) &= 0
\end{align}
The perturbation equation is rewritten as 
\begin{align}
\partial_t \theta'_\omega + \nabla \cdot \left(\bm{u}_\omega' \theta_\omega'  - \langle \bm{u}' \theta' \rangle  + \langle \bm{u}\rangle \theta_\omega' - \kappa \nabla \theta'_\omega \right) &= - \nabla \cdot \left( \bm{u}_\omega' \langle \theta \rangle \right)
\end{align}
This is an infinite system (or finite depending on the number of ensemble members) of coupled pde's between the different ensemble members. The ensemble members are coupled due to the presence of the turbulent flux, $\langle \bm{u}' \theta' \rangle$. The key observation is to notice the terms on the left hand side involve the perturbation variables and not the ensemble mean of the gradients. Assuming it is possible to find the inverse, the Green's function for the large linear system is used to yield
\begin{align}
\theta'_\omega(\bm{x}, t)  &= -\int d\bm{x}' dt' d\mu_{\alpha} \mathcal{G}_{\alpha \omega}(\bm{x}, t | \bm{x}' , t' )  \nabla \cdot \left( \bm{u}_\alpha' \langle \theta \rangle  \right) 
\end{align}
where we also have to integrate with respect to the measure defining the different ensemble members through $d \mu_{\alpha}$. Notation wise this would means $\langle \theta \rangle = \int d \mu_{\omega} \theta_\omega $. We use this expression to rewrite the turbulent flux as 
\begin{align}
\label{full_thing}
\langle \bm{u}' \theta' \rangle   &= -\int d\bm{x}' dt' d\mu_{\omega} d\mu_{\alpha} \bm{u}'_\omega(\bm{x}, t) \mathcal{G}_{\alpha \omega}(\bm{x}, t | \bm{x}' , t' ) \left[ \nabla \cdot \left( \bm{u}_\alpha'(\bm{x}' , t') \langle \theta \rangle(\bm{x}', t')  \right) \right]
\end{align}
We make two simplifications for illustrative purposes.
\begin{enumerate}
    \item All ensemble averages are independent of time.
    \item The flow is incompressible, i.e.,  $\nabla \cdot \bm{u} = 0$.
\end{enumerate}
Equation \ref{full_thing} becomes
\begin{align}
\label{partial_thing}
\langle \bm{u}' \theta' \rangle   &= -\int d\bm{x}' dt' d\mu_{\omega} d\mu_{\alpha} \left[ \bm{u}'_\omega(\bm{x}, t) \mathcal{G}_{\alpha \omega}(\bm{x}, t | \bm{x}' , t' )  \bm{u}_\alpha'(\bm{x}' , t')  \right]   \cdot \nabla \langle \theta \rangle (\bm{x}' ) 
\end{align}
We perform the $t', \alpha, \omega$ integrals first to define the turbulent-diffusivity tensor kernel as 
\begin{align}
\label{partial_thing}
\langle \bm{u}' \theta' \rangle   &= -\int d\bm{x}'  \underbrace{\int dt' d\mu_{\omega} d\mu_{\alpha} \left[ \bm{u}'_\omega(\bm{x}, t) \otimes \bm{u}_\alpha'(\bm{x}' , t')  \mathcal{G}_{\alpha \omega}(\bm{x}, t | \bm{x}' , t' )   \right]  }_{\bm{\mathcal{K}}(\bm{x}|\bm{x}')} \cdot \nabla \langle \theta \rangle (\bm{x}' ) \\
&= -\int d \bm{x}' \bm{\mathcal{K}}(\bm{x}|\bm{x}') \cdot \nabla \langle \theta \rangle(\bm{x}')
\end{align}
The independence of $\bm{\mathcal{K}}$ with respect to $t$ follows from the time-independence of $\langle \bm{u}' \theta' \rangle$ and $\langle \theta \rangle$.  In total we see 
\begin{align}
\label{closure_term}
 \langle \bm{u} \theta \rangle = \langle \bm{u} \rangle \langle \theta \rangle -  \int d \bm{x}' \bm{\mathcal{K}}(\bm{x}|\bm{x}') \cdot \nabla \langle \theta \rangle(\bm{x}').
\end{align}

An insight from \eref{closure_term} is the dependence of turbulent fluxes $\langle \bm{u}' \theta ' \rangle $ at location $\bm{x}$ as a weighted sum of gradients of the mean variable $\langle \theta \rangle$ at locations $\bm{x}'$. The operator is linear and amenable to computation, even in turbulent flows, \cite{Neeraja2020}.

We consider the spectrum for the turbulent diffusivity operator $\int d \bm{x}' \mathcal{K}(\bm{x} | \bm{x}' ) \bullet $ as a characterization of turbulent-mixing by the flow field $\bm{u}(\bm{x}, t)$. We comment that the operator $\int d \bm{x}' \mathcal{K}(\bm{x} | \bm{x}' ) \bullet $ is a mapping from vector fields to vector fields whereas the kernel $\mathcal{K}(\bm{x} | \bm{x}' )$ is a mapping from two positions to a tensor. 

For example, consider a one-dimensional problem in a periodic domain $x \in [0, 2 \pi )$. If $\mathcal{K}(x | x' ) = \kappa_e \delta(x - x')$ for some positive constant $\kappa_e$, the spectrum of the operator is flat and turbulent-mixing remains the same on every length scale. If $\mathcal{K}(x| x' ) = -\kappa_e \partial_{xx}^2 \delta(x - x')$ then the rate of mixing increases with increasing wavenumber, one gets hyperdiffusion. And lastly, if $\int dx' \mathcal{K}(x | x' ) \bullet = (\kappa_e - \partial_{xx})^{-1}$, then the kernel is nonlocal and the rate of mixing decreases at smaller length scales. 

In the following section we calculate $\int d \bm{x}' \mathcal{K}(\bm{x} | \bm{x}' ) \bullet $ directly from the conditional equations and then discuss the general structure in Section \ref{sec:general_theory}. 

\section{Examples}
\label{sec:examples_section}
We now go through three examples to understand the implications of Equations \ref{conditional_mean_equation_P_preview}-\ref{conditional_mean_equation_theta_preview}. The three examples follow sequentially in increasing complexity. The first example considers transitions between two Markov states. There we introduce a generalizable approach to computing the turbulent diffusivity. We then apply the same approach to a slightly more complex problem, transitions between three Markov states. And finally, we conclude with a calculation involving transitions between four Markov states where we delve into details with cellular flow states. The generator for the two and three state systems are derived from a finite volume discretization of an Ornstein-Uhlenbeck process, see Appendix \ref{derivation_ou}.

Although we present the general form of the kernels here, in Sections \ref{s:two_state} and \ref{s:three_state} one can consider the case where the advection-diffusion equation is one-dimensional, periodic, and with a flow field that is constant in space. In this case one can decompose the advection-diffusion equation into Fourier modes that are decoupled from one another and allows for alternative computations of the same result using more standard techniques.

\subsection{Two State}
\label{s:two_state}
To start we consider the simplest time-dependent mean-zero incompressible stochastic flow field: the transition between two incompressible states $
\bm{u}_1(\bm{x}) = \bm{u}(\bm{x}) $ and $\bm{u}_2(\bm{x}) = -\bm{u}(\bm{x})$ where each state is equally likely in the statistically steady state. For this we use the generator
\begin{align}
\label{two_state_transition}
\mathcal{Q} &=
\gamma
\begin{bmatrix}
- 1 & 1 \\
1 & -1
\end{bmatrix}
\end{align}
where $\gamma > 0$. The eigenvector and eigenvalues of the generator are
\begin{align}
\bm{v}^1 = 
\begin{bmatrix}
1/2 \\
1/2
\end{bmatrix}
\text{ and }
\bm{v}^2 = 
\begin{bmatrix}
1/2 \\
-1/2
\end{bmatrix}
\end{align}
with respective eigenvalues $\lambda^1 = 0$ and $\lambda^2 = - 2 \gamma$. The first eigenvector is the steady state probability, which we see has probability $1/2$ for each state, as expected.

In what follows we denote the Laplacian by $\Delta$. Equations \ref{conditional_mean_equation_P_preview}-\ref{conditional_mean_equation_theta_preview} for the two-state system are
\begin{align}
\partial_t \mathcal{P}_1 &= - \gamma \mathcal{P}_1 + \gamma \mathcal{P}_2 
\\
\partial_t \mathcal{P}_2 &= - \gamma \mathcal{P}_2 + \gamma \mathcal{P}_1
\\
\partial_t \Theta_1 +\nabla \cdot \left( \bm{u}  \Theta_1 \right) &= \kappa \Delta \Theta_1 + s(\bm{x}) \mathcal{P}_1 - \gamma \Theta_1 + \gamma \Theta_2 \\
\partial_t \Theta_2 -   \nabla \cdot \left(  \bm{u} \Theta_2 \right) &= \kappa \Delta \Theta_2 + s(\bm{x}) \mathcal{P}_2  - \gamma \Theta_2  + \gamma \Theta_1.
\end{align}
We assume a statistically steady state so that the temporal derivatives vanish. Consequently, $\mathcal{P}_1 = \mathcal{P}_2 = 1/2$ and the equations reduce to
\begin{align}
\nabla \cdot \left( \bm{u}  \Theta_1 \right) &= \kappa \Delta \Theta_1 + s(\bm{x}) /2 - \gamma \Theta_1 + \gamma \Theta_2 \\
- \nabla \cdot \left(  \bm{u} \Theta_2 \right) &= \kappa \Delta \Theta_2 + s(\bm{x})/2  - \gamma \Theta_2  + \gamma \Theta_1 
\end{align}

We rewrite the above equation set into a mean and perturbation by changing basis to $\varphi_i$ variables according to 
\begin{align}
\label{two_state_transformation}
\begin{bmatrix}
1/2 & 1/2 \\
1/2 & -1/2
\end{bmatrix}
\begin{bmatrix}
\varphi_1 \\
\varphi_2
\end{bmatrix}
=
\begin{bmatrix}
\Theta_1  \\
\Theta_2
\end{bmatrix}
\Leftrightarrow
\begin{bmatrix}
\varphi_1 \\
\varphi_2
\end{bmatrix}
=
\begin{bmatrix}
1 & 1 \\
1 & -1
\end{bmatrix}
\begin{bmatrix}
\Theta_1  \\
\Theta_2
\end{bmatrix}
\end{align}
yielding,
\begin{align}
\nabla \cdot \left( \bm{u} \varphi_2 \right)  &= \kappa \Delta \varphi_1 + s(\bm{x}) \\
\label{two_state_perturbation_equation}
\bm{u} \cdot \nabla \varphi_1 &= \kappa \Delta \varphi_2   - 2 \gamma \varphi_2 .
\end{align}
We used the incompressibility condition to yield the representation in Equation \ref{two_state_perturbation_equation}. Our choice of basis is no accident, we used the eigenvectors of the generator, $\mathcal{Q}$, to define the transformation in Equation \ref{two_state_transformation}. We recognize the variable $\varphi_1$ as the ensemble mean $\varphi_1 = \langle \theta \rangle$ and $\varphi_2$ as a "perturbation" variable. The turbulent flux term is  $\langle \bm{u}' \theta' \rangle =  \bm{u} \varphi_2 $. We eliminate the dependence on the perturbation variable by inverting the Helmholtz operator in Equation \ref{two_state_perturbation_equation}. In total we have the following representation of the mean equation 
\begin{align}
\nabla \cdot \left( \bm{u} (\kappa \Delta - 2 \gamma)^{-1}[ \bm{u} \cdot \nabla \langle \theta \rangle ] \right)  &= \kappa \Delta \langle \theta \rangle + s(\bm{x}) 
\end{align}
from whence we extract the turbulent diffusivity operator
\begin{align}
\label{turbulent_diffusivity_operator_two_state}
\int d \bm{x}' \mathcal{K}(\bm{x} | \bm{x}') \bullet =    \bm{u} ( 2 \gamma -  \kappa \Delta )^{-1} \bm{u} .
\end{align}

We point out a few salient features of Equation \ref{turbulent_diffusivity_operator_two_state}. The inverse Helmholtz operator, $\left(2\gamma - \kappa \Delta  \right)^{-1}$, damps high spatial frequency components of ensemble mean gradients. Thus, the operator's eigenvalues decrease as one examines increasingly fine-scale structure. Intuitively, as one examines a small-scale structure, the presence
of diffusivity leads to lower turbulent fluxes, expressing the notion that it is difficult to transport something that immediately diffuses. The second observation pertains to the presence of the eigenvalue of the generator in the Helmholtz operator. If the flow field changes rapidly, transitioning between the disparate states, then $ \gamma $ is large, and one can expect the turbulent-diffusivity to be local. In other words, the flow does not stay sufficiently long time near a coherent structure. 

In this example, the non-locality of the turbulent diffusivity is enabled by the presence of the regular diffusion operator. However, in the following example, we show that this need not be the case. 

\subsection{Three State }
\label{s:three_state}
For this example we consider a Markov process that transitions between three incompressible states $\bm{u}_1(\bm{x}) = \bm{u}(\bm{x}) $, $\bm{u}_2(\bm{x}) = 0$, and $\bm{u}_3(\bm{x}) = -\bm{u}(\bm{x})$. For this let the generator be
\begin{align}
\label{three_state_transition}
\mathcal{Q} &=
\gamma
\begin{bmatrix}
-1 & 1/2 & 0 \\
1 & -1 & 1 \\
0 & 1/2 & -1
\end{bmatrix}
\end{align}
where $\gamma > 0$. The eigenvectors of the generator are
\begin{align}
\bm{v}^1 = 
\begin{bmatrix}
1/4 \\
1/2 \\
1/4
\end{bmatrix}
\text{ , }
\bm{v}^2 = 
\begin{bmatrix}
1/2 \\
0 \\
-1/2
\end{bmatrix}
\text{ and } 
\bm{v}^3 = 
\begin{bmatrix}
1/4 \\
-1/2 \\
1/4
\end{bmatrix}
\end{align}
with respective eigenvalues $\lambda^1 = 0$, $\lambda^2 = -  \gamma$, and $\lambda^3 = - 2 \gamma$.

The statistically steady three-state manifestation of Equations \ref{conditional_mean_equation_P_preview}-\ref{conditional_mean_equation_theta_preview} are
\begin{align}
\nabla \cdot \left( \bm{u}  \Theta_1 \right) &= \kappa \Delta \Theta_1 + s(\bm{x}) /4 - \gamma \Theta_1 + \gamma \Theta_2 / 2 \\
0 &= \kappa \Delta \Theta_2 + s(\bm{x})/2  - \gamma \Theta_2  + \gamma \Theta_1 + \gamma \Theta_3 \\
- \nabla \cdot \left(  \bm{u} \Theta_3 \right) &= \kappa \Delta \Theta_3 + s(\bm{x})/4  - \gamma \Theta_3  + \gamma \Theta_2 / 2
\end{align}
Similar to before we define a transformation using the eigenvectors of the generator $\mathcal{Q}$, 
\begin{align}
\label{three_state_transformation}
\begin{bmatrix}
1/4 & 1/2 & 1/4 \\
1/2 & 0 & -1/2 \\
1/4 & -1/2 & 1/4
\end{bmatrix}
\begin{bmatrix}
\varphi_1 \\
\varphi_2 \\
\varphi_3 
\end{bmatrix}
=
\begin{bmatrix}
\Theta_1  \\
\Theta_2 \\
\Theta_3
\end{bmatrix}
\Leftrightarrow
\begin{bmatrix}
\varphi_1 \\
\varphi_2 \\
\varphi_3 
\end{bmatrix}
=
\begin{bmatrix}
1 & 1 & 1 \\
1 & 0 & -1 \\
1 & -1 & 1
\end{bmatrix}
\begin{bmatrix}
\Theta_1  \\
\Theta_2 \\
\Theta_3
\end{bmatrix}
\end{align}

The resulting equations are
\begin{align}
   \nabla \cdot \left(\bm{u} \varphi_2 \right) &= \kappa \Delta \varphi_1 + s(\bm{x}) \\
\frac{1}{2} \bm{u}\cdot \nabla \varphi_1 + \frac{1}{2} \bm{u} \cdot \nabla \varphi_3     &= \kappa \Delta \varphi_2 - \gamma \varphi_2\\
 \bm{u} \cdot \nabla \varphi_2   &= \kappa \Delta \varphi_3 - 2 \gamma \varphi_3 .
\end{align}
We again comment that $\varphi_1 = \langle \theta \rangle$ and that $\varphi_2$ and $\varphi_3$ are thought of as perturbation variables. Furthermore the turbulent flux is $\langle \bm{u}' \theta' \rangle = \bm{u} \varphi_2$. We eliminate dependence on the perturbation variables $\varphi_2$ and $\varphi_3$ by first solving for $\varphi_3$ in terms of $\varphi_2$, 
\begin{align}
\varphi_3 &= \left(\kappa \Delta - 2 \gamma   \right)^{-1} \bm{u} \cdot \nabla \varphi_2
\end{align}
and then solving for $\varphi_2$ in terms of $\varphi_1$, 
\begin{align}
\varphi_2 &= \left(\kappa \Delta -\gamma -    \frac{1}{2}\bm{u} \cdot \nabla \left( \kappa \Delta -2 \gamma  \right)^{-1} \bm{u} \cdot \nabla \right)^{-1}\frac{1}{2} \bm{u} \cdot \nabla \varphi_1
\end{align}
And finally we write our equation for the ensemble mean as
\begin{align}
\nabla \cdot \left(\bm{u} \left(\kappa \Delta -\gamma -    \frac{1}{2}\bm{u} \cdot \nabla \left( \kappa \Delta -2 \gamma  \right)^{-1} \bm{u} \cdot \nabla \right)^{-1}\frac{1}{2} \bm{u} \cdot \nabla \langle \theta \rangle  \right) &= \kappa \Delta \langle \theta \rangle + s(\bm{x} )
\end{align}
from whence we extract the turbulent diffusivity operator
\begin{align}
\int d \bm{x}' \mathcal{K}(\bm{x} | \bm{x}') \bullet &= \bm{u} \left(\gamma-\kappa \Delta +   \frac{1}{2}\bm{u} \cdot \nabla \left( \kappa \Delta -2 \gamma  \right)^{-1} \bm{u} \cdot \nabla \right)^{-1}\frac{1}{2} \bm{u} .
\end{align}
We see that, unlike the two-state system, the $\kappa \rightarrow 0$ limit retains a nonlocal feature since
\begin{align}
\int d \bm{x}' \mathcal{K}(\bm{x} | \bm{x}') \bullet \rightarrow \bm{u} \left(\gamma -   \frac{1}{4 \gamma}(\bm{u} \cdot \nabla )(\bm{u} \cdot \nabla) \right)^{-1}\frac{1}{2} \bm{u}
\end{align}
and the operator $(\bm{u}\cdot \nabla)(\bm{u} \cdot \nabla)$ can have a significant spatial structure. Similar to before, larger transition rates imply increasing local structures. In the $\kappa \rightarrow 0$ limit the non-dimensional parameter of interest is the characteristic timescale of the steady flow field, $L U^{-1}$, as it compares to the characteristic timescale for transitioning, $\gamma^{-1}$. 

For the last case, we emphasize the algebraic structure of Equations \ref{conditional_mean_equation_P_preview}-\ref{conditional_mean_equation_theta_preview} by working in detail through a four-state system.  

\subsection{Four State}
\label{four_state_example}
Here we will consider a two-dimensional velocity field motivated by \cite{FlierlMcGillicuddy}. The flow is two-dimensional, periodic with $x \in [0, 2 \pi)$, and wall-bounded with $z \in [-1, 1]$. Our Markov velocity states are defined through the stream-functions
\begin{align}
    \psi_1 &= \sin(x) \cos\left(\frac{\pi}{2}z\right) \text{, }
    \psi_2 = \cos(x) \cos\left(\frac{\pi}{2}z\right) \text{, } 
    \\
    \psi_3 &= - \sin(x) \cos\left(\frac{\pi}{2}z\right) \text{, and }
    \psi_4 = - \cos(x) \cos\left(\frac{\pi}{2}z\right),
\end{align}
and the corresponding velocity states are $\bm{u}_m = ( \partial_z \psi_m, -\partial_x \psi_m ) $. The states are simply $\pi/2$ phase shifts of one another along the periodic $x$ direction and are thought of as describing a cellular flow that randomly propagates through a channel.
The generator is given by
\begin{align}
    \mathcal{Q} &= \gamma 
    \begin{bmatrix}
    -1 & 1/2 & 0 & 1/2 \\
    1/2 & -1 & 1/2 & 0 \\
    0 & 1/2 & -1 & 1/2 \\
    1/2 & 0 & 1/2 & -1
    \end{bmatrix}
    + 
    \omega 
    \begin{bmatrix}
    0 & 1/2 & 0 & -1/2 \\
    -1/2 & 0 & 1/2 & 0 \\
    0 & -1/2 & 0 & 1/2 \\
    1/2 & 0 & -1/2 & 0
    \end{bmatrix}.
\end{align}
The parameter $\gamma$ is, as in the previous examples, related to the amount of time spent in a given state. When $\omega = 0$ the cells move with equal likelihood to the left or the right, whereas $\omega \neq 0$ biases the cells to move in a particular direction. We require $|\omega | \leq \gamma$ for a probabilistic interpretation of results. 

We assume that the system is in a statistically steady state. Thus we start by observing that the normalized right eigenvectors of $\mathcal{Q}$ are
\begin{align}
\bm{v}_1 &= 
\begin{bmatrix}
 1/4 \\ 
 1/4 \\ 
 1/4 \\ 
 1/4 
\end{bmatrix}
\text{ , }
\bm{v}_2 = 
\begin{bmatrix}
  1/2  \\
  \iota /2 \\
 -1/2   \\ 
  -\iota /2   
\end{bmatrix}
\text{ , }
\bm{v}_3 = 
\begin{bmatrix}
  1/2  \\
  -\iota / 2 \\
 -1/2    \\ 
  \iota  / 2 
\end{bmatrix}
\text{ , and }
\bm{v}_4 = 
\begin{bmatrix}
  1/4  \\
  - 1/4 \\
 1/4    \\ 
  - 1/4  
\end{bmatrix}
\end{align}
with corresponding eigenvectors $\lambda_1 =  0, \lambda_2 = -\gamma - \omega \iota, \lambda_3 =  -\gamma + \omega \iota, \lambda_4 =  -2 \gamma$, respectively. Inspection of $\bm{v}^1$, the vector associated with the statistically steady state, reveals that each state is equally likely in the steady state.

The steady-state equations are 
\begin{align}
 \bm{u}_1 \cdot \nabla \Theta_1 
 &= \kappa \Delta \Theta_1 + s/4 - \gamma \Theta_1 + \frac{\gamma + \omega}{2}  \Theta_2  +  \frac{\gamma - \omega}{2}  \Theta_4 
 \\
 \bm{u}_2 \cdot \nabla \Theta_2 
 &= \kappa \Delta \Theta_2 + s/4 - \gamma \Theta_2 + \frac{\gamma + \omega}{2}  \Theta_3  + \frac{\gamma - \omega}{2}  \Theta_1 
 \\
 \bm{u}_3 \cdot \nabla \Theta_3 
 &= \kappa \Delta \Theta_3 + s/4 - \gamma \Theta_3 + \frac{\gamma + \omega}{2} \Theta_4   + \frac{\gamma - \omega}{2} \Theta_2  
 \\
\bm{u}_4 \cdot \nabla \Theta_4 
 &= \kappa \Delta \Theta_4 + s/4 - \gamma \Theta_4 + \frac{\gamma + \omega}{2} \Theta_1   + \frac{\gamma - \omega}{2}\Theta_3 .
\end{align}

As was done in prior sections we will change basis by utilizing the eigenvectors of $\mathcal{Q}$. We avoid the use of imaginary eigenvectors by instead using linear combinations $(v^2 + v^3)/2$ and $(v^2 - v^3)/(2 \iota)$ at the cost of a non-diagonal generator in the resulting basis. Explicitly we use the similarity transformation defined by the matrices
\begin{align}
S = 
\begin{bmatrix}
 1/4  &  1/2  & 0     &  1/4 \\ 
 1/4  &  0    & 1/2   & -1/4 \\
 1/4  & -1/2  & 0     &  1/4 \\
 1/4  &  0    & -1/2  & -1/4 
\end{bmatrix}
\text{ and }
S^{-1}  = 
\begin{bmatrix}
 1.0  &  1.0   & 1.0  &  1.0 \\ 
 1.0  &  0.0   & -1.0  & 0.0 \\
 0.0  & 1.0   & 0.0  &  -1.0 \\
 1.0  &  -1.0 & 1.0  & -1.0 
\end{bmatrix}
.
\end{align}
Thus we make the change of variables 
\begin{align}
 \begin{bmatrix}
 1.0  &  1.0   & 1.0  &  1.0 \\ 
 1.0  &  0.0   & -1.0  & 0.0 \\
 0.0  & 1.0   & 0.0  &  -1.0 \\
 1.0  &  -1.0 & 1.0  & -1.0 
\end{bmatrix}
\begin{bmatrix}
\Theta_1 \\
\Theta_2 \\
\Theta_3 \\
\Theta_4
\end{bmatrix}
= 
\begin{bmatrix}
\varphi_1 \\
\varphi_2 \\
\varphi_3 \\
\varphi_4
\end{bmatrix}
.
\end{align}
We observe $\varphi_1 = \langle \theta \rangle$. 
With this change of variable and using $\bm{u}_1 = - \bm{u}_3$ with $\bm{u}_2 = - \bm{u}_4$,  the resulting system of equations is written in block operator form as, 
\begin{align}
\label{block_operator_form}
\begin{bmatrix}
- \kappa \Delta  & \nabla \cdot \left( \bm{u}_1 \bullet \right) & \nabla \cdot \left( \bm{u}_2 \bullet \right)  & 0 
\\
0  & \gamma - \kappa \Delta  & -\omega & \frac{1}{2}\bm{u}_1 \cdot \nabla  
\\
0 & \omega  & \gamma - \kappa \Delta  & - \frac{1}{2} \bm{u}_2 \cdot \nabla  
\\
0    & \bm{u}_1 \cdot \nabla  & - \bm{u}_2 \cdot \nabla  & 2\gamma - \kappa \Delta 
\end{bmatrix}
\begin{bmatrix}
\varphi_1 \\
\varphi_2 \\
\varphi_3 \\
\varphi_4
\end{bmatrix}
&=
\begin{bmatrix}
s(x,z) \\
- \frac{1}{2}\bm{u}_1 \cdot \nabla \varphi_1 \\
- \frac{1}{2}\bm{u}_2 \cdot \nabla \varphi_1 \\
0 
\end{bmatrix}
\end{align}
where the dependence of the perturbation variables on $\varphi_1$ is included as a source rather than as a part of the block operator. 
The inverse of the $3 \times 3$ lower right  submatrix matrix of Equation \ref{block_operator_form},
\begin{align}
\label{greens_function_for_conditional}
\mathcal{G} = 
\begin{bmatrix}
 \gamma - \kappa \Delta  & -\omega & \frac{1}{2}\bm{u}_1 \cdot \nabla  
\\
 \omega  & \gamma - \kappa \Delta  & - \frac{1}{2} \bm{u}_2 \cdot \nabla  
\\
 \bm{u}_1 \cdot \nabla  & - \bm{u}_2 \cdot \nabla  & 2\gamma - \kappa \Delta 
\end{bmatrix}
^{-1}
\end{align}
is the Green's function associated with the perturbation variables $\varphi_2$, $\varphi_3$, $\varphi_4$. The perturbation Green's function is used to represent the turbulent diffusivity operator as
\begin{align}
\label{full_turbulent_flux}
\int d \bm{x}' \mathcal{K}(\bm{x} | \bm{x}') \bullet &= 
\begin{bmatrix}
\bm{u}_1  & 
\bm{u}_2   & 
0 
\end{bmatrix}   
        \begin{bmatrix}
 \gamma - \kappa \Delta  & -\omega & \frac{1}{2}\bm{u}_1 \cdot \nabla  
\\
 \omega  & \gamma - \kappa \Delta  & - \frac{1}{2} \bm{u}_2 \cdot \nabla  
\\
 \bm{u}_1 \cdot \nabla  & - \bm{u}_2 \cdot \nabla  & 2\gamma - \kappa \Delta 
\end{bmatrix}
^{-1}
\begin{bmatrix}
\bm{u}_1 / 2 \\
\bm{u}_2  /2 \\
0 \\
\end{bmatrix}
\end{align}
Written this way, we emphasize the Schur-complement structure of the ensemble mean equations when eliminating dependence on the perturbation variables.

To better understand the turbulent diffusivity operator for this example, we consider two local approximations. Assuming a scale separation we approximate
\begin{align}
\int  \mathcal{K}(\bm{x} | \bm{x}') \cdot \nabla \langle \theta \rangle(\bm{x}') d\bm{x}' \approx \int  \mathcal{K}(\bm{x} | \bm{x}')  d\bm{x}' \cdot \nabla \langle \theta \rangle(\bm{x}) = \bm{D}_1 \cdot \nabla \langle \theta \rangle(\bm{x}) 
\end{align}
where we integrate the turbulent diffusivity kernel with respect to $\bm{x}'$
\begin{align}
    \bm{D}_1 &\equiv \int  \mathcal{K}(\bm{x} | \bm{x}') d\bm{x}'  .
\end{align}
We comment that, upon discretization, the linear operator $\int d \bm{x}' \mathcal{K}(\bm{x} | \bm{x}') \bullet $ is represented as a matrix. Performing the integration with respect to $d \bm{x}'$ amounts to performing a row sum on the matrix.  

A second local estimate of the turbulent diffusivity is obtained by neglecting the dissipation terms and perturbation gradients, i.e.
\begin{align}
\label{particular_local_example}
    \begin{bmatrix}
 \gamma - \kappa \Delta & -\omega & \frac{1}{2}\bm{u}_1 \cdot \nabla  
\\
 \omega  & \gamma - \kappa \Delta & - \frac{1}{2} \bm{u}_2 \cdot \nabla  
\\
 \bm{u}_1 \cdot \nabla  & - \bm{u}_2 \cdot \nabla  & 2\gamma - \kappa \Delta
\end{bmatrix}
^{-1}
\approx 
\begin{bmatrix}
\gamma  & -\omega & 0 
\\
\omega  & \gamma  & 0 
\\
0 & 0  & 2\gamma
\end{bmatrix}
^{-1}
.
\end{align}
Thus the local eddy diffusivity is 
\begin{align}
\label{second_loca_diffusivity}
  \bm{D}_2 &= 
\begin{bmatrix}
\bm{u}_1  & 
\bm{u}_2   & 
0 
\end{bmatrix}   
\begin{bmatrix}
\gamma  & -\omega & 0 
\\
\omega  & \gamma  & 0 
\\
0 & 0  & 2\gamma
\end{bmatrix}
^{-1}
\begin{bmatrix}
\bm{u}_1 / 2 \\
\bm{u}_2  /2 \\
0 \\
\end{bmatrix}
\\
&= 
\frac{1}{2(\gamma^2 + \omega^2)} \left[ \gamma \left(\bm{u}_1 \otimes \bm{u}_1 + \bm{u}_2 \otimes \bm{u}_2 \right) + \omega \left(\bm{u}_2 \otimes \bm{u}_1 - \bm{u}_1 \otimes \bm{u}_2 \right) \right]
\end{align}
where we have interpreted products of vector fields as outer products. Using the velocity fields, 
\begin{align}
    \bm{u}_1 &= -\frac{\pi}{2} \sin(x) \sin \left(\frac{\pi}{2} z \right) \hat{x} - \cos(x) \cos \left( \frac{\pi}{2} z \right) \hat{z} 
    \\
    \bm{u}_2 &= -\frac{\pi}{2} \cos(x) \sin \left(\frac{\pi}{2} z \right) \hat{x} + \sin(x) \cos \left( \frac{\pi}{2} z \right) \hat{z},
\end{align}
we compute each outer product to obtain the components of the local turbulent diffusivity,
\begin{align}
[\bm{D}_2]_{\hat{x} \otimes \hat{x}}
    &=
    \frac{\gamma \pi^2}{8(\gamma^2 + \omega^2)} \sin^2 \left(\frac{\pi}{2} z \right)
    \\
[\bm{D}_2]_{\hat{x} \otimes \hat{z}} &=  -[\bm{D}_2]_{\hat{z} \otimes \hat{x}} =         \frac{\omega \pi}{4(\gamma^2 + \omega^2)} 
    \sin \left( \frac{\pi}{2} z \right) \cos \left( \frac{\pi}{2} z \right)
    \\
[\bm{D}_2]_{\hat{z} \otimes \hat{z}} 
    &=
    \frac{\gamma }{2(\gamma^2 + \omega^2)} \cos^2 \left(\frac{\pi}{2} z \right)
\end{align}
In Section \ref{sec:general_theory} we show, in general, the equivalence of  neglecting dissipation terms and perturbation gradients, as was done in Equation \ref{particular_local_example}, and  estimating the diffusivity by computing the integrated auto-correlation of the statistically steady velocity field,
\begin{align}
\bm{D}_2 = \int_0^\infty \langle \bm{u}(\bm{x}, t + \tau ) \otimes \bm{u} ( \bm{x}, t ) \rangle d \tau .
\end{align}

\begin{figure}
  \centerline{\includegraphics[width=0.75\textwidth]{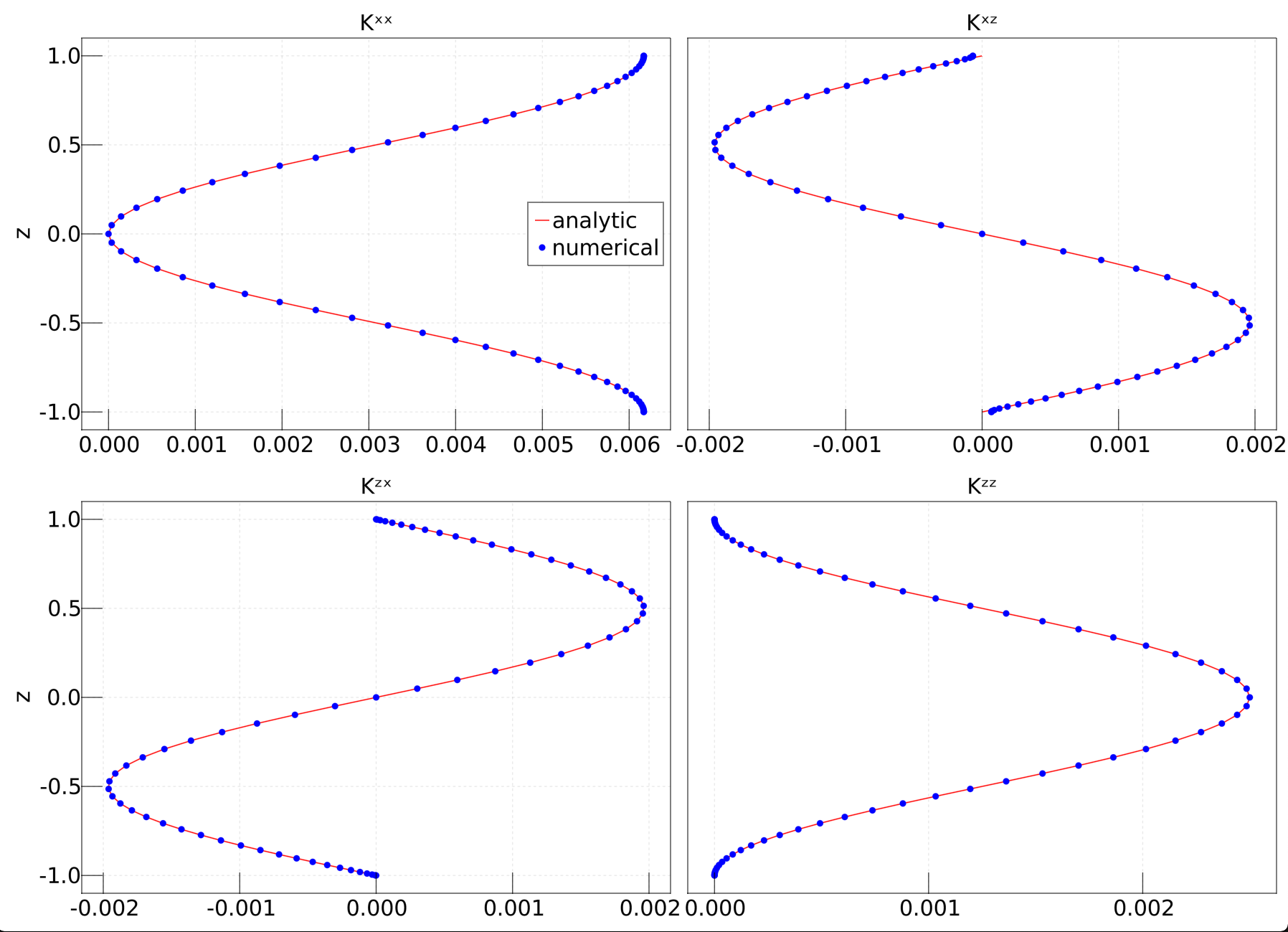}}
  \caption{A comparison of two local diffusivity estimates. The analytic diffusivity estimate is in red and numerically computed diffusivity estimate uses blue dots. The z-axis is depth and the x-axis is the diffusivity amplitude. Here $\gamma = \omega = 100$ and $\kappa = 1/100$. }
\label{fig:local_local_diffusivity}
\end{figure}

\begin{figure}
  \centerline{\includegraphics[width=0.75\textwidth]{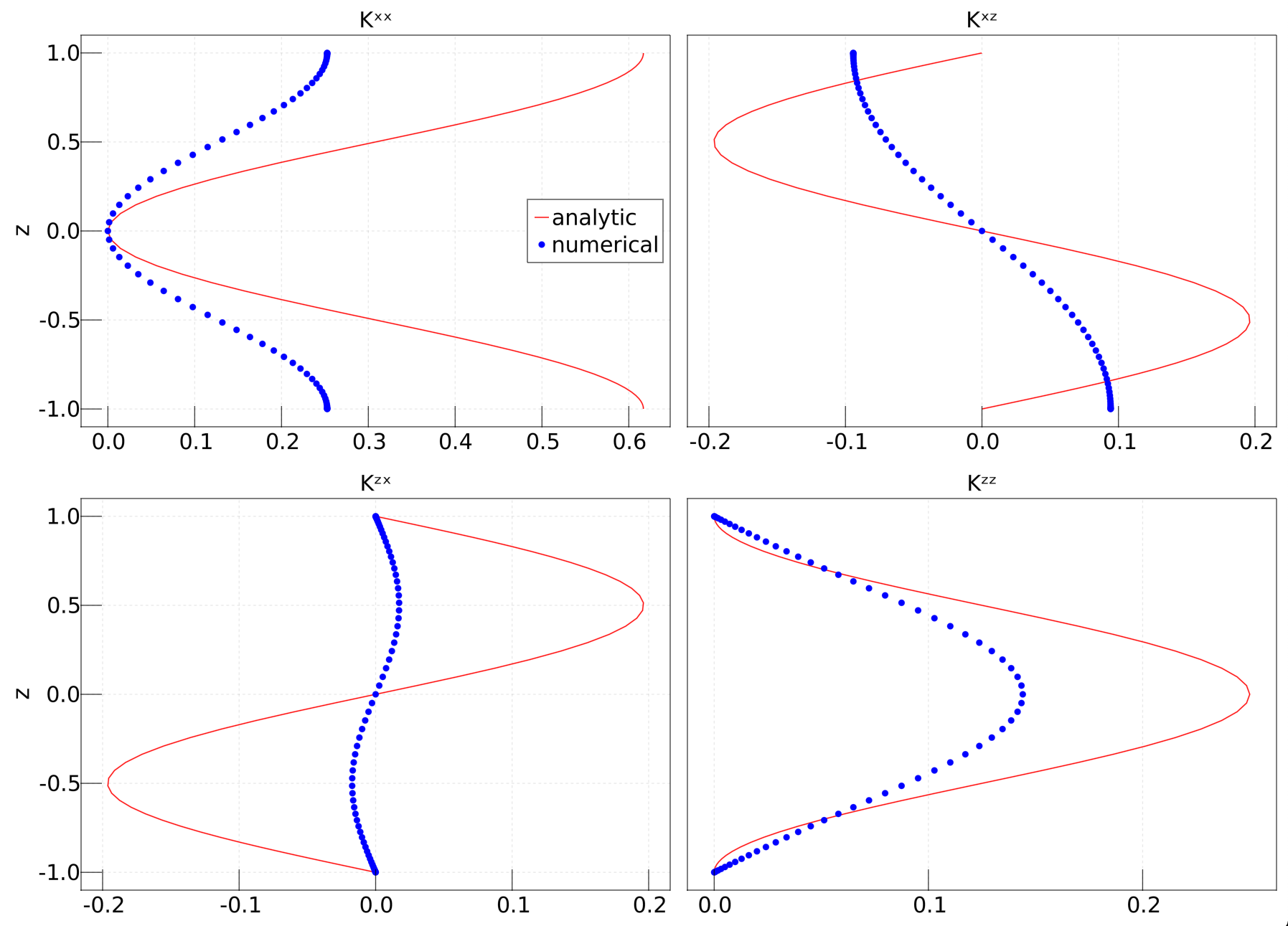}}
  \caption{A comparison of two local diffusivity estimates. The analytic diffusivity estimate is in red and numerically computed diffusivity estimate uses blue dots. The z-axis is depth and the x-axis is the diffusivity amplitude. Here $\gamma = \omega = 1$ and $\kappa = 1$.}
\label{fig:nonlocal_local_diffusivity}
\end{figure}

\begin{figure}
  \centerline{\includegraphics[width=1\textwidth]{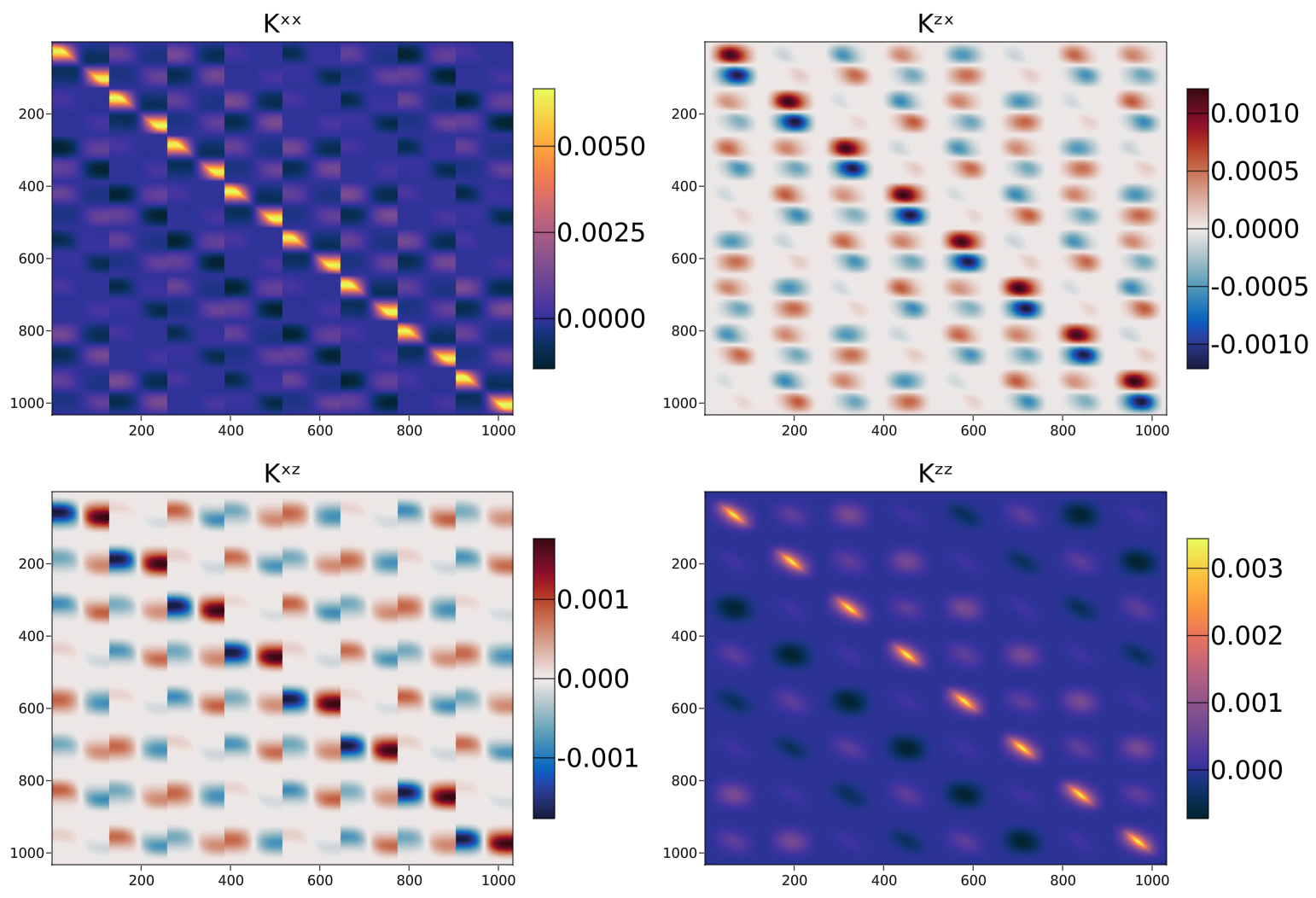}}
  \caption{A representation of the turbulent diffusivity operator corresponding to the case in Figure \ref{fig:nonlocal_local_diffusivity}. If the diffusivity was indeed local then each of the form matrices would only have a diagonal component. The x-z coordinate axis have been collapsed to a single index, hence the banded structure of the output. Thus the rows are the "output" axis corresponding to the $(x,z)$ of $\int dx' dz' K(x,z|x',z') \bullet $ and the columns are the $(x',z')$ corresponding to input. The ordering is chosen such that each block-diagonal structure corresponds to a fixed $x_i, x_j'$ grid location. The similarity between different block rows follow from the periodicity of the $x$ coordinate.  Here $\gamma = \omega = 1$ and $\kappa = 1$.} 
\label{fig:nonlocal_kernel_diffusivity}
\end{figure}

We illustrate the local diffusivity calculation in two scenarios: 
\begin{enumerate}
    \item $\gamma = \omega = 100$ and $\kappa = 1/100$
    \item $\gamma = \omega = \kappa = 1$
\end{enumerate}
In the first case, we expect the local diffusivity estimate to work well, and the two diffusivity estimates to correspond to one another. In the latter case, there is no scale separation between transition rates, diffusive timescales, and advective timescales; we expect nonlocal effects to play a significant role. We discretize all operators using a collocation method as described by \cite{SpectralMethodsTrefethen} to explore the nonlocality of the full turbulent flux operator, Equation \ref{full_turbulent_flux}. We use 65 Chebyshev modes in the wall-bounded direction and 8 Fourier modes in the periodic direction to approximate each operator, leading to a $1560 \times 1560$ sized matrix representation of Equation \ref{greens_function_for_conditional}. There are no significant changes upon halving or doubling the resolution in each direction. Furthermore, given the independence of the periodic direction in $\bm{D}_2$, we use the local estimate 
\begin{align}
    \overline{\bm{D}}_1(z) &=  \frac{1}{2\pi}\int  \mathcal{K}(x, z | x', z') dx' dz' dx
\end{align}
for comparison between $\bm{D}_1$ and $\bm{D}_2$. 

For the first case, we compute the local diffusivities and display their result in Figure \ref{fig:local_local_diffusivity}. We see that there is excellent agreement between the two approaches, except near the boundary in the $K^{xz}$ component of the diffusivity tensor. Here we note the influence of the homogenous Neumann boundary conditions in the diffusivity estimate. 

For the second case, we compute the local diffusivities and display their result in Figure \ref{fig:local_local_diffusivity}. We see that there is poor agreement between the two approaches in each component of the diffusivity tensor. Given the lack of scale separation between the timescales of the problem, this comes as no surprise.

To further explore the discrepancy, we show the full operator, $\int dx' dz' \mathcal{K}(x,z | x', z') \bullet$, in Figure \ref{fig:nonlocal_kernel_diffusivity}. The current turbulent diffusivity operator is a four-dimensional object, which we represent with a two-dimensional heatmap. To do so, we flatten both the $x$ and $z$ dimensions into a single index and order them so that the $z$ values are sequential and $x$ values are separated upon the completion of a $z$-range. The structured pattern of the heatmap  in Figure \ref{fig:nonlocal_kernel_diffusivity} is a consequence of the periodicity of the $x$ direction. There is a block structure in each of the components of the operator. These blocks correspond to the diffusivity operator at a fixed $x, x'$ location and represent the variation in $z, z'$. Thus the diagonal block component corresponds to the $x=x'$ part of the operator. Each block row corresponds to a fixed $x$ value, and each block column corresponds to a fixed $x'$ value. One can count 8 blocks appearing on a given row (and column), which corresponds to our choice of using 8 Fourier modes in the periodic direction. Each block row seems to be a periodic translation of one another; however, we emphasize that this is merely in appearance rather than actuality. The four states are not sufficient to guarantee translation invariance. 

The rich structure of the current operator stands in stark contrast to a local diffusivity operator. A local diffusivity operator is a diagonal matrix. We emphasize that the significant off-diagonal components imply that turbulent fluxes are not related solely to local gradients but must incorporate a weighted sum of gradients in a neighborhood of a given location.

In the following section, we gather the approach used in the examples and generalize.

\section{General Approach}
\label{sec:general_theory}
We have seen three examples that all follow a similar pattern: 
\begin{enumerate}
    \item Compute the eigenvectors of the generator $\mathcal{Q}$.
    \item Transform the equations into a basis that diagonalizes $\mathcal{Q}$.
    \item Separate the mean equation from the perturbation equations.
    \item Solve for the perturbation variables in terms of the mean variable.
\end{enumerate}
Here we aim to gather the above procedure in the general case where we have access to the eigenvectors of $\mathcal{Q}$.
Furthermore, in the last example, we claimed that the local turbulent diffusivity approximation as calculated by neglecting the effects of diffusion and perturbation gradients is equivalent to calculating the integrated auto-correlation of the Markov process. We justify that claim in Section \ref{local_approximation}.

\subsection{Notation}
Let us establish a notation for the general procedure. We again let $\mathcal{Q}$ denote the generator with corresponding transition probability matrix $\mathscr{P}(\tau)$ given by the matrix exponential
\begin{align}
\mathscr{P}(\tau) &= \exp(\tau \mathcal{Q}).
\end{align}
The entries of the matrix $[\mathscr{P}(\tau)]_{mn}$ denotes the transition probability of state $n$ to the state $m$. In each column of the transition matrix the sum of the entries is one. We assume a unique zero eigenvalue for $\mathcal{Q}$ with all other eigenvalues negative. We also assume that the eigenvalues can be ordered in such a way that they are decreasing, i.e. $\lambda_1 = 0$, $\lambda_2 < 0$ and $\lambda_i \leq \lambda_j$ for $i > j$ with $j \geq 2$. These choices result in a unique statistical steady state which we denote by the vector $\bm{v}_1$ with the property
\begin{align}
\mathcal{Q} \bm{v}_1 &= 0 \bm{v}_1 \text{ and } \mathscr{P}(\tau) \bm{v}_1 = \bm{v}_1 \text{ for all } \tau .
\end{align}
and similarly for the left eigenvector, $\bm{w}_1$.
We denote the entries of $\bm{v}_1$ and $\bm{w}_1$ by column vectors
\begin{align}
\bm{v}_1 &= 
\begin{bmatrix}
P_1 \\
P_2 \\
\vdots \\
P_M
\end{bmatrix}
\text{ and }
\bm{w}_1 = 
\begin{bmatrix}
1 \\
1 \\
\vdots \\
1
\end{bmatrix}
\end{align}
where $M$ is the number of states. We assume that the eigenvector $\bm{v}_1$ is normalized such that $\sum_m P_m = 1$. Consequently, $\bm{w}_1 \cdot \bm{v}_1 = \bm{w}_1^T \bm{v}_1 = 1$. We introduce unit vectors $\hat{\bm{e}}_m$ whose $m'th$ entry is zero and all other entries are zero, e.g.
\begin{align}
    \hat{\bm{e}}_1
    &=
    \begin{bmatrix}
1 \\
0 \\
\vdots \\
0
\end{bmatrix}
\text{ , }
    \hat{\bm{e}}_2
    =
    \begin{bmatrix}
0 \\
1 \\
\vdots \\
0
\end{bmatrix}
\text{ and }
\hat{\bm{e}}_M
=
\begin{bmatrix}
0 \\
0 \\
\vdots \\
1
\end{bmatrix}
.
\end{align}
Thus, $\bm{v}_1 = \sum_m P_m \bm{\hat{e}}_m$, $\bm{w}_1 = \bm{\hat{e}}_m $. Furthermore, $\bm{\hat{e}}_m \cdot \bm{v}_1 = P_m$ and $\bm{\hat{e}}_m \cdot \bm{w}_1 = 1$ for each $m$.

For the discussion that follows we will assume that the matrix $\mathcal{Q}$ has an eigenvalue decomposition. In general we denote the right eigenvectors of $\mathcal{Q}$ by $\bm{v}_i$ for $i = 1, .., M$ and the left eigenvectors by $\bm{w}_i$ for $i = 1, ..., M$. These vectors are all associated with eigenvalues $\lambda_i$ for $i=1, ..., M$ where $i=1$ denotes the unique eigenvalue $\lambda_1 = 0$. We recall that the left eigenvectors can be constructed from the right eigenvectors by stacking all the left eigenvectors in a matrix $V$, computing the inverse $V^{-1}$, and extracting the rows of the inverse. The aforementioned procedure guarantees the normalization $\bm{w}_j \cdot \bm{v}_i =  \bm{w}_i^T \bm{v}_j = \delta_{ij}$. Thus we have the relations
\begin{align}
 \mathcal{Q} \bm{v}_n = \lambda_n \bm{v}_n  \text{ and }   \bm{w}^T_n \mathcal{Q} = \lambda_n \bm{w}^T_n  .
\end{align}

With notation now in place, we observe that the operators $\mathcal{Q}$ and $\mathscr{P}(\tau)$ are characterized by their spectral decomposition
\begin{align}
    \mathcal{Q} &= \sum_i \lambda_i \bm{v}_i \bm{w}_i^T
    \text{ and }
    \mathscr{P}(\tau) = \sum_i e^{\tau \lambda_i} \bm{v}_i \bm{w}_i^T.
\end{align}
We remind the reader of the various use of "P"s and their relation:
\begin{enumerate}
    \item $\mathbb{P}$ denotes a probability. 
    \item $\mathscr{P}(\tau)$ denotes the transition probability matrix for a time $\tau$ in the future.
    \item $\mathcal{P}_m(t)$ denotes the probability of being in state $m$ at time $t$. The algebraic relation \[ \sum_m P_m(t+\tau) \bm{\hat{e}}_m = \mathscr{P}(\tau) \sum_n P_n(t) \bm{\hat{e}}_n \] holds. 
    \item $P_m$ is the statistically steady probability of being found in state $m$. In the limit \[ \lim_{t \rightarrow \infty} \mathcal{P}_m(t) = P_m . \] 
\end{enumerate}

We now introduce our Markov states as steady vector fields. The use of several vector spaces imposes a burden on notation: The vector spaces associated with Markov states, ensemble members, and the vector field $\bm{u}$. Instead of using overly decorated notation with an excessive number of indices, we introduce the convention that $\bm{u}$ will always belong to the vector space associated with the vector field, and all other vectors are associated with the vector space of Markov states. Effectively we let elements of our vector space associated with Markov states belong to a different algebra than real numbers.

\subsection{The Spectral Representation}
With this notation now in place, the statistically steady equations 
\begin{align}
 \nabla \cdot \left( \bm{u}_m  \Theta_m  \right)&=   \kappa \Delta \Theta_m + P_m s  + \sum_{n} \mathcal{Q}_{mn} \Theta_n
\end{align}
are represented as the matrix system
\begin{align}
\label{conditional_mean_in_vector_form}
    \sum_{m} \hat{\bm{e}}_m \nabla \cdot \left( \bm{u}_m  \Theta_m  \right)&= \sum_{m} \hat{\bm{e}}_m  \kappa \Delta \Theta_m + s \bm{v}_1 +   \mathcal{Q} \left(\sum_{m} \hat{\bm{e}}_m  \Theta_m \right)
\end{align}
where we made use of $\sum_m \bm{\hat{e}}_m P_m = \bm{v}_1$. We now re-express Equation \ref{conditional_mean_in_vector_form} in terms of a basis that uses the eigenvectors of the transition matrix. Define components $\varphi_n$ by the change of basis formula 
\begin{align}
    \sum_m \Theta_m \bm{\hat{e}}_m &=  \sum_n \varphi_n \bm{v}_n \Leftrightarrow  \sum_n \varphi_n \bm{\hat{e}}_n = \sum_{mn} (\bm{w}_n \cdot \bm{\hat{e}}_m)\bm{\hat{e}}_n \Theta_m 
\end{align}
We make the observation $\varphi_1 = \langle \theta \rangle$. We have the following relations based on the general definitions of the left eigenvectors $\bm{w}_n$ and right eigenvectors $\bm{v}_n$, 
\begin{align}
\label{change_of_basis_relations}
\Theta_n  &= \sum_i (\bm{\hat{e}}_n \cdot \bm{v}_i ) \varphi_i  \text{ and } \varphi_n = \sum_m (\bm{w}_n \cdot \bm{\hat{e}}_m) \Theta_m .
\end{align} 
Multiplying Equation \ref{conditional_mean_in_vector_form} by $\bm{w}^T_j$ and making use of Equation \ref{change_of_basis_relations} we get
\begin{align}
\label{conditional_mean_in_vector_form}
\sum_{n} (\bm{w}_j \cdot \hat{\bm{e}}_n) \nabla \cdot \left( \bm{u}_n \Theta_n \right)   &= \kappa \Delta \varphi_j + \delta_{1j} s +  \lambda_j \varphi_j 
\\
\nonumber
&\Rightarrow 
\\
\sum_{n} (\bm{w}_j \cdot \hat{\bm{e}}_n) \nabla \cdot \left( \bm{u}_n \left[ \sum_i \bm{\hat{e}}_n \cdot \bm{v}_i \varphi_i \right] \right) &= \kappa \Delta \varphi_j + \delta_{1j} s +  \lambda_j \varphi_j  
\\
\nonumber
&\Rightarrow 
\\
\label{conditional_mean_in_spectral_form}
\sum_{in} (\bm{w}_j \cdot \hat{\bm{e}}_n)(\bm{\hat{e}}_n \cdot \bm{v}_i ) \nabla  \cdot \left( \bm{u}_n \varphi_i  \right) &= \kappa \Delta \varphi_j + \delta_{1j} s +  \lambda_j \varphi_j
\end{align}
We now wish to decompose Equation \ref{conditional_mean_in_spectral_form} into a mean equation, index $j=1$, and perturbation equations $j>1$. For the mean equation, we make use of the properties
\begin{align}
    \lambda_1 = 0 \text{ , } \bm{w}_1 \cdot \hat{\bm{e}}_n = 1 \text{ , and } \sum_{(i=1)n} (\bm{w}_1 \cdot \hat{\bm{e}}_n)(\bm{\hat{e}}_n \cdot \bm{v}_i )\nabla \cdot \left( \bm{u}_n \varphi_i  \right) = \nabla \cdot \langle  \bm{u}   \varphi_1 \rangle 
\end{align}
to arrive at (after changing summation index from $n$ to $m$),
\begin{align}
\label{mean_equations_from_spectral}
\nabla \cdot \left( \langle \bm{u} \rangle \varphi_1 \right) + \nabla \cdot \left[ \sum_{(i\neq 1) m} (\bm{\hat{e}}_m \cdot \bm{v}_i ) \bm{u}_m \varphi_i  \right] &= \kappa \Delta \varphi_1 + s
\end{align}
We make the observation that the turbulent flux is
\begin{align}
    \langle \bm{u}' \theta' \rangle &= \sum_{(i\neq 1) m} (\bm{\hat{e}}_m \cdot \bm{v}_i ) \bm{u}_m \varphi_i  .
\end{align}
The perturbation equations, indices $j > 1$, are
\begin{align}
\label{perturbation_equations_from_spectral}
\sum_{in} (\bm{w}_j \cdot \hat{\bm{e}}_n)(\bm{\hat{e}}_n \cdot \bm{v}_i ) \nabla \cdot \left( \bm{u}_n  \varphi_i  \right) &= \kappa \Delta \varphi_j +  \lambda_j \varphi_j   \text{ for } j > 1 .
\end{align}
We isolate the dependence on the mean gradients by rearranging the above expression as follows for $j > 1$
\begin{align}
\label{schur_preperation}
\sum_{(i \neq 1) n} (\bm{w}_j \cdot \hat{\bm{e}}_n)(\bm{\hat{e}}_n \cdot \bm{v}_i ) \nabla \cdot  \left( \bm{u}_n  \varphi_i  \right) - \kappa \Delta \varphi_j -  \lambda_j \varphi_j  &= - \sum_n  P_n (\bm{w}_j \cdot \hat{\bm{e}}_n) \nabla \cdot \left( \bm{u}_n  \varphi_1  \right) 
\end{align}
where we used $\bm{\hat{e}}_n \cdot \bm{v}_i  = P_n$. Assuming that the operator on the left-hand side of Equation  \ref{schur_preperation} is invertible, we introduce the Green's function, $\mathcal{G}_{ij}$ to yield
\begin{align}
\varphi_i = - \int d \bm{x}' \sum_{(j \neq 1) n} \mathcal{G}_{ij}(\bm{x} | \bm{x}') P_n (\bm{w}_j \cdot \hat{\bm{e}}_n) \nabla_{\bm{x}'} \cdot \left( \bm{u}_n(\bm{x}')  \varphi_1(\bm{x}')  \right) \text{ for } i \neq 1
\end{align}
Thus we represent our turbulent flux as
\begin{align}
    \langle \bm{u}' \theta' \rangle &= - \int d \bm{x}'  \sum_{(i\neq 1)(j \neq 1) m n } (\bm{\hat{e}}_m \cdot \bm{v}_i ) \bm{u}_m (\bm{x}) \mathcal{G}_{ij}(\bm{x} | \bm{x}') P_n (\bm{w}_j \cdot \hat{\bm{e}}_n) \nabla_{\bm{x}'} \cdot \left( \bm{u}_n(\bm{x}')  \varphi_1(\bm{x}')  \right)   .
\end{align}
For compressible flow, the eddy-flux depends on both the ensemble mean gradients and the ensemble mean value; otherwise, when each Markov state is incompressible,
\begin{align}
    \langle \bm{u}' \theta' \rangle &= - \int d \bm{x}'  \sum_{(i\neq 1)(j \neq 1) m n  } (\bm{\hat{e}}_m \cdot \bm{v}_i ) \bm{u}_m (\bm{x}) \mathcal{G}_{ij}(\bm{x} | \bm{x}') P_n (\bm{w}_j \cdot \hat{\bm{e}}_n) \bm{u}_n(\bm{x}')  \cdot \nabla_{\bm{x}'}   \varphi_1(\bm{x}'), 
\end{align}
in which case the turbulent diffusivity kernel is 
\begin{align}
\mathcal{K}(\bm{x} | \bm{x}') &= \sum_{(i\neq 1) (j \neq 1) m n  } (\bm{\hat{e}}_m \cdot \bm{v}_i ) \bm{u}_m (\bm{x}) \mathcal{G}_{ij}(\bm{x} | \bm{x}') P_n (\bm{w}_j \cdot \hat{\bm{e}}_n) \bm{u}_n(\bm{x}') . 
\end{align}
The above expression completes the procedure that we enacted for the examples in Section \ref{sec:examples_section}. 

We now discuss local approximations to the turbulent diffusivity operator. 

\subsection{Local Approximation}
\label{local_approximation}
We start with the same local diffusivity approximation of Section \ref{four_state_example} but using the spectral representation of Equations \ref{conditional_mean_equation_P_preview}-\ref{conditional_mean_equation_theta_preview}. In the perturbation equations, neglect the dissipation operator and perturbation gradients, e.g. only include index $i=1$, to yield the following reduction of Equation \ref{perturbation_equations_from_spectral},
\begin{align}
\label{taylored_perturbation_equations_from_spectral}
\sum_{n} (\bm{w}_i \cdot \hat{\bm{e}}_n) P_n \left( \bm{u}_n \cdot \nabla \varphi_1  \right) &=  \lambda_i \varphi_i   \text{ for } i > 1
\end{align}
where we used $(\bm{\hat{e}}_n \cdot \bm{v}_1 ) = P_n$ and have changed indices from $j$ to $i$. We solve for $\varphi_i$ for $i  > 1$ and focus on the perturbation flux term in Equation \ref{mean_equations_from_spectral} 
\begin{align}
\langle \bm{u}' \theta' \rangle = \sum_{(i\neq 1) m} (\bm{\hat{e}}_m \cdot \bm{v}_i ) \bm{u}_m \varphi_i
\end{align}
to get the local turbulent-diffusivity estimate,
\begin{align}
\label{taylored_hypothesis}
\sum_{(i\neq 1) m} (\bm{\hat{e}}_m \cdot \bm{v}_i ) \bm{u}_m \varphi_i &= \sum_{(i\neq 1) m} (\bm{\hat{e}}_m \cdot \bm{v}_i ) \bm{u}_m \left[\frac{1}{\lambda_i} \sum_{n} (\bm{w}_i \cdot \hat{\bm{e}}_n) P_n \left( \bm{u}_n \cdot \nabla \varphi_1  \right) \right] \\
\label{taylored_hypothesis_2}
&= \underbrace{\left[ \sum_{(i \neq 1)mn} \frac{-1}{\lambda_i} ( \bm{\hat{e}}_m \cdot \bm{v}_i  ) (\bm{w}_i \cdot \bm{\hat{e}}_n)  \bm{u}_m  \otimes P_n \bm{u}_n \right]}_{\bm{D}} \cdot (- \nabla \varphi_1) .
\end{align}

We aim to show that the turbulent diffusivity from Equation \ref{taylored_hypothesis_2} 
\begin{align}
\label{previous_estimate}
\bm{D} &= \sum_{(i \neq 1)mn} \frac{-1}{\lambda_i} ( \bm{\hat{e}}_m \cdot \bm{v}_i  ) (\bm{w}_i \cdot \bm{\hat{e}}_n)  \bm{u}_m  \otimes P_n \bm{u}_n 
\end{align}
is equivalent to estimating the diffusivity by calculating the integral of the velocity perturbation autocorrelation in a statistically steady state,
\begin{align}
\bm{D} = \int_0^\infty \langle \bm{u}'(\bm{x},t + \tau) \otimes \bm{u}'(\bm{x}, t ) \rangle d\tau
\end{align}
The above turbulent diffusivity is expected to work well in the limit that diffusive effects can be neglected and the velocity field transitions rapidly with respect to the advective timescale. Under such circumstances it is not unreasonable to think of velocity fluctuations as analogous to white noise with a given covariance structure. For example, letting $\bm{\xi}$ be a white noise process and $\bm{\sigma}$ be a variance vector, if
\begin{align}
    \bm{u}'(\bm{x},t) \approx \bm{\sigma}(\bm{x}) \xi \text{ where } \langle \xi(t+ \tau) \xi(t ) \rangle = \delta(\tau)
\end{align}
then a diffusivity is given by
\begin{align}
    \bm{D}(\bm{x}) = \int_0^\infty \langle \bm{u}'(\bm{x},t + \tau) \otimes \bm{u}'(\bm{x}, t ) \rangle d\tau &=  \bm{\sigma}(\bm{x}) \otimes \bm{\sigma}(\bm{x}) .
\end{align}
Indeed, we will show that the intuitive estimate,
\begin{align}
\label{local_diffusivity_correlation_estimate}
    \bm{D}(\bm{x}) = \int_0^\infty \langle \bm{u}'(\bm{x},t + \tau) \otimes \bm{u}'(\bm{x}, t ) \rangle d\tau
\end{align}
does correspond to Equation \ref{previous_estimate}.

We begin with two observations. First, the statistically steady velocity field satisfies
\begin{align}
\langle \bm{u}(\bm{x}, t) \rangle = \sum_m P_m \bm{u}_m(\bm{x} ),
\end{align}
where $\bm{u}_m(\bm{x})$ for each $m$ are the states of the Markov process.  Second, recall that the vector $\mathscr{P}(\tau) \bm{\hat{e}}_n $ is a column vector of probabilities whose entries denote the probability of being found in state $m$ given that at time $\tau = 0$ the probability of being found in state $n$ is one. Thus, the conditional expectation of $\bm{u}(\bm{x}, t + \tau )$ given $\bm{u}(\bm{x}, t ) = \bm{u}_n(\bm{x})$ is 
\begin{align}
\label{conditional_expectation}
    \langle \bm{u}(\bm{x}, t + \tau ) \rangle_{\bm{u}(\bm{x}, t ) = \bm{u}_n(\bm{x}) } &= \left(\sum_m \bm{u}_m (\bm{x}) \hat{\bm{e}}_m \right)^T \mathscr{P}(\tau) \bm{\hat{e}}_n \\
    &= \sum_{im} e^{\tau \lambda_i} ( \bm{\hat{e}}_m \cdot \bm{v}_i  ) (\bm{w}_i \cdot \bm{\hat{e}}_n) \bm{u}_m (\bm{x}).
\end{align}
Equation \ref{conditional_expectation} expresses the conditional expectation as a weighted sum of Markov states $\bm{u}_m(\bm{x})$.

We are now in a position to characterize the local turbulent-diffusivity estimate. The local turbulent-diffusivity is computed by taking the long time integral of a statistically steady flow field's autocorrelation function, i.e.
\begin{align}
\bm{D}(\bm{x}) = \int_0^\infty \bm{R}(\bm{x}, \tau) d \tau 
\end{align}
where 
\begin{align}
\label{autocorrelation_definition}
\bm{R}(\bm{x}, \tau) \equiv  \langle \bm{u}(\bm{x}, t + \tau)  \otimes \bm{u}(\bm{x}, t) \rangle - \langle
 \bm{u}(\bm{x}, t + \tau) \rangle \otimes \langle\bm{u}(\bm{x}, t) \rangle . 
\end{align}
We calculate the second term under the statistically steady assumption of Equation \ref{autocorrelation_definition},
\begin{align}
\label{other_term}
 \langle
 \bm{u}(\bm{x}, t + \tau) \rangle \otimes \langle\bm{u}(\bm{x}, t) \rangle  = \left( \sum_m P_m \bm{u}_m(\bm{x} ) \right) \otimes \left( \sum_n P_n \bm{u}_n(\bm{x} ) \right) .
\end{align}
For the first term of Equation \ref{autocorrelation_definition} we decompose the expectation into conditional expectations,
\begin{align}
\label{decomposition}
\langle \bm{u}(\bm{x}, t + \tau)  \otimes \bm{u}(\bm{x}, t)  \rangle = \sum_n \langle \bm{u}(\bm{x}, t + \tau)  \otimes \bm{u}(\bm{x}, t)  \rangle_{\bm{u}(\bm{x}, t ) = \bm{u}_n(\bm{x}) }P_n
\end{align}
Given that we are in a statistically steady state, we use Equation \ref{conditional_expectation} to establish
\begin{align}
\langle \bm{u}(\bm{x}, t + \tau)  \otimes \bm{u}(\bm{x}, t)  \rangle &= \sum_n \langle \bm{u}(\bm{x}, t + \tau)  \otimes \bm{u}(\bm{x}, t)  \rangle_{\bm{u}(\bm{x}, t ) = \bm{u}_n(\bm{x}) } P_n 
\\
&= \sum_{imn} e^{\tau \lambda_i} ( \bm{\hat{e}}_m \cdot \bm{v}_i  ) (\bm{w}_i \cdot \bm{\hat{e}}_n) \bm{u}_m \otimes  P_n \bm{u}_n.
\end{align}
We isolate the $i=1$ index and use $\lambda_1 = 0$, $\bm{\hat{e}}_m \cdot \bm{v}_1  = P_m$, and $\bm{w}_1 \cdot \bm{\hat{e}}_n = 1$  to arrive at
\begin{align}
\label{zerothmode}
\sum_{mn} e^{\tau \lambda_1} ( \bm{\hat{e}}_m \cdot \bm{v}_1  ) (\bm{w}_1 \cdot \bm{\hat{e}}_n)   \bm{u}_m \otimes  P_n \bm{u}_n = \left(\sum_m P_m \bm{u}_m \right)  \otimes \left(\sum_n P_n \bm{u}_n \right) .
\end{align}
Equation \ref{zerothmode} cancels with \ref{other_term} so that in total we have the following characterization of Equation \ref{autocorrelation_definition} 
\begin{align} 
\label{autocorrelation_characterization}
\bm{R}(\bm{x}, \tau) &= \sum_{(i \neq 1)mn} e^{\tau \lambda_i} ( \bm{\hat{e}}_m \cdot \bm{v}_i  ) (\bm{w}_i \cdot \bm{\hat{e}}_n)  \bm{u}_m  \otimes P_n \bm{u}_n.
\end{align}
Equation \ref{autocorrelation_characterization} is integrated to yield the local turbulent-diffusivity
\begin{align}
\label{cute_formula}
\bm{D}(\bm{x}) = \int_0^\infty \bm{R}(\bm{x}, \tau) d \tau = \sum_{(i \neq 1)mn} \frac{-1}{\lambda_i} ( \bm{\hat{e}}_m \cdot \bm{v}_i  ) (\bm{w}_i \cdot \bm{\hat{e}}_n)  \bm{u}_m  \otimes P_n \bm{u}_n
\end{align}
where we used $\lambda_i < 0$ for $i > 1$. A comparison of Equation \ref{cute_formula} to Equation \ref{previous_estimate} reveals the correspondence. Thus we see that estimating the diffusivity through the velocity autocorrelation integral is equivalent to neglecting diffusive effects and perturbation gradients.

\section{Conclusions}
We have introduced a class of stochastic partial differential equations amenable to analysis in this work. The class of problems falls under the umbrella of stochastic advection, where the flow state is modeled as a continuous time Markov process. We reformulated the problem of finding a turbulence closure for passive scalars advected by a stochastic flow field into solving a set of partial differential equations by conditionally averaging the passive scalar equation with respect to the flow state.

The resulting dimensionality of the equations depended on the number of variables required to describe flow statistics and the dimensionality of the flow. A flow characterized by $m$ discrete variables leads to a set of $m$-coupled equations of the same dimensionality as the original. A flow characterized by a continuum of statistical variables can be discretized and reduced to the former. Eliminating the system's dependence on all but the ensemble mean leads to an operator characterization of the turbulence closure, allowing for an exploration of closures that don't invoke a scale separation hypothesis.

We explored three examples of increasing complexity--Markov states characterized by two, three, and four states--and outlined a general approach to obtaining a closure based on the spectrum of the transition probability operator. In the examples, we examined the role of non-locality in determining a statistically steady turbulence closure. We calculated closures for all three systems and numerically evaluated a Green's function for the four-state system. We also found the small velocity amplitude, weak scalar diffusivity, and fast transition rate limit reduce the closure to a spatially heterogeneous tensor acting on ensemble mean gradients. Furthermore, we related this tensor to the time-integrated auto-correlation of the stochastic flow field. 

We have not exhausted the number of examples offered by the formulation nor simplifications leading to analytically tractable results. Interesting future directions include using Markov states estimated directly from turbulence simulations, analyzing scale-separated flows, generalizing the advection-diffusion equation to reaction-advection-diffusion equations, and formulating optimal mixing problems. When the number of Markov states increases, the computational burden of estimating turbulent diffusivity operators becomes demanding; thus, there is a need to develop methods that exploit the structure of the problem as much as possible. 

Mathematically there are many challenges as well. All the arguments provided here are formal calculations, and the necessity of rigorous proofs remains. For example, a direct proof of the conditional averaging procedure is necessary. Ultimately, the goal is to reduce the stochastic-advection turbulence closure problem to one that can leverage theory from partial differential equations.

\backsection[Supplementary data]{\label{SupMat}Supplementary material and movies are available at 
\\ ZENODO 
\\ \url{https://github.com/sandreza/StatisticalNonlocality}}

\backsection[Acknowledgements]{We would like to thank the 2018 Geophysical Fluid Dynamics summer school where much of this work was completed. We would also like to thank Tobias Bischoff, Simon Byrne, and Raffaele Ferrari for their encouragement and discussion with regards to the present manuscript. }

\backsection[Funding]{Our work is supported by the generosity of Eric and Wendy Schmidt by recommendation of the Schmidt Futures program, and by the National Science Foundation under grant AGS-6939393.  }

\backsection[Declaration of interests]{The authors report no conflict of interest.}

\appendix

\section{An Alternative Formal Derivation}
\label{FeynmanStyle}
We wish to show that one can work directly with the continuous formulation of the advection-diffusion equations for the derivation of the conditional mean equations. Although we consider a finite (but arbitrarily large) number of Markov states here, considering a continuum follows mutatis mutandi. In Section \ref{sec:theory} we wrote down the master equation for the discretized stochastic system as 
    \begin{align}
\label{discrete_master_equation}
    \partial_t \rho_m &= \sum_{i} \frac{\partial}{\partial \theta^i} \left[ \left( \sum_{jkc} A_{ijk}^c u^{k,c}_m \theta^j  - \sum_j D_{ij} \theta^j -  s^i  \right ) \rho_m \right] + \sum_n \mathcal{Q}_{mn} \rho_n .
\end{align}
We introduce the (spatial) volume element $\Delta \bm{x}_i$ to rewrite Equation \ref{discrete_master_equation} in the evocative manner,
\begin{align}
\label{discrete_master_equation_prep}
    \partial_t \rho_m &= \sum_{i} \Delta \bm{x}_i \frac{1}{\Delta \bm{x}_i} \frac{\partial}{\partial \theta^i} \left[ \left( \sum_{jkc} A_{ijk}^c u^{k,c}_m \theta^j  - \sum_j D_{ij} \theta^j -  s^i  \right ) \rho_m \right] + \sum_n \mathcal{Q}_{mn} \rho_n .
\end{align}
We now take limits
\begin{align}
    \sum_{i}\Delta \bm{x}_i &\overset{``\text{lim}"}{ = } \int d \bm{x}   , 
    \\
    \frac{1}{\Delta \bm{x}_i } \frac{\partial}{\partial \theta^i } &\overset{``\text{lim}"}{ = } \frac{\delta}{\delta \theta (\bm{x} )} ,
    \\
    \sum_{jkc} A_{ijk}^c u^{k,c}_m \theta^j  + \sum_j D_{ij} \theta^j -  s^i  
    &\overset{``\text{lim}"}{ = } 
    \bm{u}_m \cdot \nabla \theta - \kappa \Delta \theta - s ,
\end{align}
to get the functional evolution equation for the probability density,
\begin{align}
\label{continuous_fokker_planck}
    \partial_t \rho_m &= \int d \bm{x} \frac{\delta}{\delta \theta (\bm{x} )} \left( \left[ \bm{u}_m \cdot \nabla \theta - \kappa \Delta \theta - s \right] \rho_m  \right) + \sum_n \mathcal{Q}_{mn} \rho_n
\end{align}
where $\bm{x}$ is a continuous index. As before we can derive the CM equations directly from the above equations. To do so we make the additional correspondence 
\begin{align}
d \bm{\theta} &\overset{``\text{lim}"}{ = } \mathcal{D}[\theta] .
\end{align}
We now define the same quantities as before, but using the field integral
\begin{align}
\mathcal{P}_m &\equiv \int \mathcal{D}[\theta] \rho_m \text{ and } \Theta_m(\bm{y}) \equiv \int \mathcal{D}[\theta] \theta( \bm{y} )\rho_m .
\end{align}
We mention that the discrete indices $i,j,k$ from Equations \ref{sec:theory} before get replaced by the continuous labels such as $\bm{x}$ and $\bm{y}$. We only make use of a few formal properties of the field integral, with direct correspondence the the $n$-dimensional integrals. We use linearity, i.e. for two mappings with compatible ranges $\mathcal{F}[\theta]$ and $\mathcal{H}[\theta]$,
\begin{align}
    \int \mathcal{D}[ \theta] \left( \mathcal{F}[\theta] + \mathcal{H}[\theta]\right) &= \int \mathcal{D}[\theta ] \mathcal{F}[\theta]  + \int \mathcal{D}[\theta] \mathcal{H}[\theta] 
\end{align}
We use the analogue to the divergence theorem, 
\begin{align}
\int \mathcal{D}[\theta] \int d \bm{x} \frac{\delta}{\delta \theta (\bm{x} )} \left( \left[ \bm{u}_m \cdot \nabla \theta - \kappa \Delta \theta - s \right] \rho_m  \right) &= 0
\\
    \nonumber 
    &\Leftrightarrow 
    \\
    \int d\bm{\theta} \nabla_{\bm{\theta}} \cdot ( \bm{f} \rho) &= 0
\end{align}
since the integral of a divergence should be zero if the probabilities vanish at infinity (i.e. that our tracer cannot have infinite values at a given point in space). We also make use of the integration by parts, i.e. for some functionals $\mathcal{F}$ and $\mathcal{H}$,
\begin{align}
   \int \mathcal{D}[\theta] \mathcal{H} \int d \bm{x} \frac{\delta}{\delta \theta (\bm{x} )} \mathcal{F} &= -  \int \mathcal{D}[\theta]  \int d \bm{x} \frac{\delta \mathcal{H} }{\delta \theta (\bm{x} )}  \mathcal{F} 
   \\
   \nonumber 
   &\Leftrightarrow 
   \\
   \int d\bm{\theta} h \nabla_{\bm{\theta}} \cdot \bm{f} &= - \int d \bm{\theta} \left( \nabla_{\bm{\theta}} h  \right) \cdot \bm{f}
\end{align}
And finally we also interchange sums and integrals, 
\begin{align}
    \int \mathcal{D}[\theta ] (\Delta \theta) \rho_m &= \Delta \int \mathcal{D}[\theta] \theta \rho_m = \Delta \Theta_m  \\
    \nonumber 
    &\Leftrightarrow
    \\
    \int d \bm{\theta} \left( \sum_j D_{\ell j} \theta^j \rho_m \right) &= \sum_j D_{\ell j} \int d \bm{\theta} \theta^j \rho_m = \sum_j D_{\ell j} \Theta^j_m .
\end{align}
We proceed similarly for the $\bm{u}_m \cdot \nabla$ term. We also use properties of the variational derivative such as, 
\begin{align}
    \frac{\delta \theta(\bm{y})}{\delta \theta (\bm{x})} = \delta (\bm{x} - \bm{y}) \Leftrightarrow \frac{\partial \theta^\ell}{\partial \theta^i} = \delta_{\ell i} .
\end{align}
Taken together one can directly obtain Equations \ref{conditional_mean_equation_P_preview} and \ref{conditional_mean_equation_theta_preview} by first integrating Equation \ref{continuous_fokker_planck} with respect to $\mathcal{D}[\theta] $ to get
\begin{align}
    \partial_t \mathcal{P}_m &= \sum_n \mathcal{Q}_{mn} \mathcal{P}_n
\end{align}
and multiplying Equation \ref{continuous_fokker_planck} by $\theta(\bm{y})$ and then integrating with respect to $\mathcal{D}[\theta]$ to get 
\begin{align}
\partial_t \Theta_m(\bm{y}, t) + \nabla_{\bm{y}} \cdot \left( \bm{u}_m(\bm{y}) \Theta_m(\bm{y}, t) - \kappa \nabla_{\bm{y}} \Theta_m(\bm{y}, t) \right) &= s(\bm{y}) \mathcal{P}_m  + \sum_n \mathcal{Q}_{mn} \Theta_n(\bm{y}, t) .
\end{align}
In the above expression, removing explicit dependence of the position variable yields, 
\begin{align}
\partial_t \Theta_m + \nabla \cdot \left( \bm{u}_m \Theta_m - \kappa \nabla \Theta_m \right) &= s P_m  + \sum_n \mathcal{Q}_{mn} \Theta_n .
\end{align}

Our reason for mentioning the above methodology is that it allows for expedited computations. There is no need to explicitly discretize, perform usual $n$-dimensional integral manipulations, and then take limits afterward. For example, computing the conditional two-moment equations defined by the variable
\begin{align}
C_m(\bm{y}, \bm{z}, t) &\equiv \int \mathcal{D}[\theta] \theta(\bm{y}) \theta(\bm{z}) \rho_m ,
\end{align}
is obtained by multiplying Equation \ref{continuous_fokker_planck} by $\theta(\bm{y})$ and $\theta(\bm{z})$ and integrating with respect to $\mathcal{D}[\theta]$,
\begin{align}
\partial_t C_m + \nabla_{\bm{y}} \cdot \left( \bm{u}_m (\bm{y}) C_m - \kappa \nabla_{\bm{y}} C_m \right) + \nabla_{\bm{z}} \cdot \left( \bm{u}_m(\bm{z}) C_m - \kappa \nabla_{\bm{z}} C_m \right) \\
= s(\bm{z}) \Theta_m(\bm{y})  + s(\bm{y}) \Theta_m(\bm{z}) + \sum_{n}\mathcal{Q}_{mn} C_n .
\end{align}
In particular we note the source term on the right hand side and the appearance of the first conditional moment. In the derivation we used the product rule
\begin{align}
    \frac{\delta ( \theta(\bm{y}) \theta(\bm{z}) )}{\delta \theta(\bm{x}) } &= \delta( \bm{x} - \bm{y} ) \theta(\bm{z} ) + \delta(\bm{x} - \bm{z}) \theta ( \bm{y} ) .
\end{align}
If the advection-diffusion equation is $m$-dimensional and we have $N$ Markov states, the above equation is $2m + N$ dimensional. Indeed the equation for the $M'th$ moment is a $M \times m + N$ dimensional partial differential equation.
\section{A Heuristic Overview of the Master Equation and Discretizations}
\label{heuristic_overview}
In this section we provide an argument for the form of the master equation in the main text, Equation \ref{discrete_master_equation} in Section \ref{sec:theory}. Our starting point is Section \ref{two_var} where we use the Liouville equation for two continuous variables. We then apply the finite volume method to the Fokker-Planck equation of an Ornstein-Uhlenbeck process to derive the transition matrices used in the two-state and three-state systems in Section \ref{sec:examples_section}. We conclude with a formal argument for the use of discrete Markov states as an approximation to the compressible Euler equations in \ref{shady_bit}.

\subsection{Two Variable System}
\label{two_var}
Suppose that we have an two variables $x, y \in \mathbb{R}$ governed by the equations,
\begin{align}
\frac{dx}{dt} &= f(x) + \sqrt{2} \sigma \xi  \\
\frac{dy}{dt} &= g(x,y).
\end{align}
where $\xi$ is white noise. 
In this context we think of $x$ as being our flow field $\bm{u}$ and $y$ as the tracer $\theta$. The master equation implied by the dynamics is
\begin{align}
\label{liouville_xy}
\partial_t \rho &= - \partial_x \left(f(x) \rho - \sigma^2 \partial_x \rho \right)  - \partial_y \left( g(x,y) \rho \right).
\end{align}
We now discretize the equation with respect to the $x-$variable by partitioning space into non-overlapping cells, characterized by domains $\Omega_m$. First we start with the Fokker-Planck equation for $x$, which is independent of the $y-$variable,
\begin{align}
\label{liouville_x}
\partial_t P &= - \partial_x \left(f(x) P - \sigma^2 \partial_x P\right) .
\end{align}
Observe the relation $\int \rho(x,y,t) dy = P(x,t)$. Define our coarse grained variable $\mathcal{P}_m$ as 
\begin{align}
\mathcal{P}_m &\equiv \int_{\Omega_m} P(x) dx 
\end{align}
which is a probability. Thus the discretization of Equation \ref{liouville_x} becomes  
\begin{align}
\label{master_x}
\partial_t \int_{\Omega_m} P dx &= - \int_{\Omega_m} \partial_x \left(f(x) \rho - \sigma^2 \partial_x  \rho \right) dx  
\\
\nonumber
&\approx
\\
\label{master_x_done}
\partial_t \mathcal{P}_m &= \sum_n \mathcal{Q}_{mn} \mathcal{P}_n
\end{align}
for a generator $\mathcal{Q}$ which we derive in \ref{derivation_ou} with respect to a chosen numerical flux. Heuristically, going from Equation \ref{master_x}-\ref{master_x_done} is accomplished by observing that $\mathcal{P}_m$ is a probability and the operator $\mathcal{L} \equiv  \partial_x \left(f(x) \bullet - \sigma^2 \partial_x \bullet \right)$ is linear; thus, upon discretization, the operator is represented a matrix\footnote{It is, of course, possible to approximate using a nonlinear operator, but for simplicity we only consider the linear case.} acting on the chosen coarse grained variables $\mathcal{P}_n$. The property $\sum_m \mathcal{Q}_{mn} = \bm{0}$ is the discrete conservation of probability.

Going back to Equation \ref{liouville_xy}, defining
\begin{align}
\rho_m(y) &\equiv \int_{\Omega_m} \rho(x,y) dx, 
\end{align}
introducing $x_m \in \Omega_m$,
and performing the same discretization for the joint Markov system yields,
\begin{align}
\label{master_xy}
\partial_t \rho_m &= \sum_n \mathcal{Q}_{mn} \rho_n - \partial_y \left( g(x_m, y) \rho_m \right),
\end{align}
where we used the approximation
\begin{align}
\int_{\Omega_m} g(x,y) \rho(x,y) dx &\approx   g(x_m, y) \int_{\Omega_m} \rho(x,y)dx  =  g(x_m,y) \rho_m(y).
\end{align}
The $x_m$ are the Markov states and the $\mathcal{Q}_{mn}$ serves as the specification for transitioning between different states. We also observe that one can simply start with the discrete states for $x$ and continuous variables for $y$ to directly obtain Equation \ref{master_xy} as was done in the main text. 

In what follows we give a concrete example of deriving a transition matrix $\mathcal{Q}$ from a finite-volume discretization of an Ornstein-Uhlenbeck (OU) process. We explicitly mention the kind of discretization that we use since retaining mimetic properties of the transition matrix $\mathcal{Q}$ is not guaranteed with other discretizations. Furthermore, using a finite volume discretization allows for the resulting discretization to be interpreted as a continuous time Markov process with a finite state space. 

\subsection{Example Discretization}
\label{derivation_ou}
Consider an Ornstein-Uhlenbeck process and the resulting Fokker-Planck equation,
\begin{align}
    \partial_t \rho &= - \partial_x \left(- x \rho - \partial_x \rho\right).
\end{align}
We discretize the above equation with $N+1$ cells, where $N = 1$ and $N = 2$ correspond to the two and three state systems, respectively. Using a finite volume discretization, we take our cells to be
\begin{align}
\Omega_m &=   [ \Delta x \left(m - 1/2 - N/2\right) , \Delta x \left( m + 1/2 - N/2 \right)]  \\
\Delta x &= \frac{2}{\sqrt{N}}
\end{align}
for $m = 0, 1, ..., N$. Our choice implies that cell centers (the discrete Markov states) are 
\begin{align}
\label{cell_centers}
    x_m &= \Delta x(m - N/2)
\end{align}
for $m = 0 , ..., N$ and cell faces are
\begin{align}
x_m^f &= \Delta x (m - 1/2 - N/2)   
\end{align}
for $m = 0, ..., N+1$. 
We define
\begin{align}
 \mathcal{P}_m &=  \int_{\Omega_m } \rho dx \text{ and } \overline{\rho}_m \Delta x = \mathcal{P}_m.
\end{align}
Upon integrating with respect to the control volume we obtain,
\begin{align}
\frac{d}{dt} \mathcal{P}_m &= -\left.  \left(- x \rho - \partial_x \rho\right) \right|_{x = x_m^f} + \left. \left(- x \rho - \partial_x \rho\right) \right|_{x = x_{m+1}^f}
\end{align}
The numerical flux is chosen as follows,
\begin{align}
\label{discretization_ou_flux}
\left. \left(- x \rho - \partial_x \rho \right) \right|_{x = x_{m}^f} &\approx 
-\frac{ x_{m-1} \overline{\rho}_{m-1} + x_{m} \overline{\rho}_m }{2} - \frac{\overline{\rho}_{m} - \overline{\rho}_{m-1}}{\Delta x} 
\\
&= -\frac{ x_{m-1} \mathcal{P}_{m-1} + x_{m} \mathcal{P}_m }{2 \Delta x} - \frac{\mathcal{P}_{m} - \mathcal{P}_{m-1}}{(\Delta x)^2} \\
&= \frac{1}{2}\left( (N-m+1) \mathcal{P}_{m-1} -m \mathcal{P}_m \right)
\end{align}
where we use the convention $\mathcal{P}_{-1} = \mathcal{P}_{N+1} = 0$ so that boundaries, corresponding to indices $m=0$ and $m=N+1$, imply no flux conditions. 
Combining the flux estimates for both cell boundaries, the evolution equation for the probabilities $\mathcal{P}_m$ becomes
\begin{align}
\label{equation_row}
    \partial_t \mathcal{P}_m &=  \frac{1}{2}\left[ ( N-m +1 ) \mathcal{P}_{m-1} - N \mathcal{P}_{m} + (m+1) \mathcal{P}_{m+1} \right],
\end{align} 
which implies the transition matrix
\begin{align}
\label{general_transition}
    \mathcal{Q}_{mn} &= \frac{1}{2}\left( -N \delta_{mn} + n \delta_{(m+1)n} + (N-n)\delta_{(m-1)n} \right).
\end{align}
Equation \ref{equation_row} emphasize the row structure of the matrix whereas Equation \ref{general_transition} emphasizes the column structure. 
The steady state probability distribution is the binomial distribution\footnote{The continuous steady state distribution is a Normal distribution $\rho(x) = (2\pi)^{-1/2} \exp \left( -x^2/2 \right)$.} 
\begin{align}
    P_m &= 2^{-N} \binom{N}{m}.
\end{align}
Furthermore, the eigenvectors and eigenvalues of the matrix are in correspondence with the eigenfunctions and eigenvalues of the OU process as noted by \cite{Doering_1989}. In particular, the cell centers, Equation \ref{cell_centers}, is a left eigenvector of $\mathcal{Q}_{mn}$ with eigenvalue $\lambda = -1$. This relation is useful for calculating the auto-correlation of the Markov process since Equation \ref{conditional_expectation} only involves one eigenvalue.  We used the transition matrix, Equation  \ref{general_transition}, in the construction of the two and three state systems.

Similar to the construction in this section, four-state system transition matrix is obtained from discretizing a random walk in a periodic domain with a drift. The term proportional to $\gamma$ is attributed to diffusion and the term proportional to $\omega$ is attributed to drift. The resulting cell centers are then taken as the phase in the periodic direction of a fixed stream function.

\subsection{A Finite Volume Approximation in Function Space}
\label{shady_bit}
We start with the compressible Euler-Equations
\begin{align}
    \partial_t \rho + \nabla \cdot \left(\rho \bm{u} \right) &= 0 ,
    \\
    \partial_t \rho \bm{u} + \nabla \cdot \left( \rho \bm{u} \otimes \bm{u} \right) + \nabla p &= 0 ,
    \\
    \partial_t \rho e + \nabla \cdot \left( \bm{u} \left[ \rho e + p \right]  \right)
    &= 0 ,
    \\
    p(\rho, \rho \bm{u}, \rho e )  &= p
\end{align}
where $\rho$ is density, $\rho \bm{u}$ is the momentum, $\bm{u} = \rho \bm{u} / \rho$ is the velocity, $\rho e$ is the total energy density, and $p$ is a thermodynamic pressure\footnote{For example, one could use the pressure for an ideal gas $p = (\gamma - 1) (\rho e - \rho |\bm{u}|^2 /2)$ with $\gamma = 7/5$.}. Here we introduce $Z$ as a probability density in function space for the state variables $S \equiv (\rho, \rho \bm{u} , \rho e)$. In the notation of \ref{FeynmanStyle}, the evolution equation for the statistics $Z$ are 
\begin{align}
\label{shady_probability_density}
 \partial_t Z &= \int d\bm{x} \left[
 \frac{\delta}{\delta \rho}\left( \nabla \cdot \left[\rho \bm{u} \right] Z \right) 
 + \frac{\delta}{\delta \rho \bm{u} }\left( \nabla \cdot \left[ \rho \bm{u} \otimes \bm{u} \right] Z + \nabla p Z \right) + \frac{\delta}{ \delta \rho e} \left( \nabla \cdot \left( \bm{u} \left[ \rho e + p \right]  \right) Z \right) \right] ,
\end{align}
where we have suppressed the index $\bm{x}$ in the variational derivatives.

Now consider a partition in function space into domains $\Omega_m$ and let $S_m$ denote a value of a state within the set $S_m$. In this case we define the probability as 
\begin{align}
\mathcal{P}_m &\equiv \int_{\Omega_m} \mathcal{D}[\rho] \mathcal{D}[\rho \bm{u}] \mathcal{D}[ \rho e] Z .
\end{align}
In analogy with the calculations in Section \ref{derivation_ou}, integrating equation \ref{shady_probability_density} with respect to a control volume $\Omega_m$ would result in an equation of the form
\begin{align}
\partial_t \mathcal{P}_m = \sum_n  \mathcal{Q}_{mn}\mathcal{P}_n
\end{align}
for some generator $\mathcal{Q}_{mn}$. The entries of the generator are functionals of the states $S_m \in \Omega_m$. Performing the necessary integrals and re-expressing it in this finite form is done indirectly through data-driven methods with time-series as in \cite{Klus2016}, \cite{Fernex2021}, or \cite{Schumacher2022}. The difficulty of performing a discretization from first principles comes from choosing the subsets of function space to partition and carrying out the integrals in function space. Periodic orbits and fixed points of a flow serve as a natural skeleton for function space, but are typically burdensome to compute. We offer no solution, but hope that in the future such direct calculations are rendered tractable. In the meanwhile, indirect data-driven methods are the most promising avenue for the calculation of the generator $\mathcal{Q}$. 

\bibliographystyle{jfm}
\bibliography{references}

\begin{thebibliography}{23}
\expandafter\ifx\csname natexlab\endcsname\relax\def\natexlab#1{#1}\fi
\def\au#1{#1} \def\ed#1{#1} \def\yr#1{#1}\def\at#1{#1}\def\jt#1{\textit{#1}}
  \def\bt#1{#1}\def\bvol#1{\textbf{#1}} \def\vol#1{#1} \def\pg#1{#1}
  \def\publ#1{#1}\def\arxiv#1{#1}\def\org#1{#1}\def\st#1{\textit{#1}}

\bibitem[Allawala \& Marston(2016)]{Marston2016}
{\sc \au{Allawala, Altan} \& \au{Marston, J.~B.}} \yr{2016}  \at{Statistics of
  the stochastically forced lorenz attractor by the fokker-planck equation and
  cumulant expansions}.  \jt{Phys. Rev. E}  \bvol{94},  \pg{052218}.

\bibitem[{Avellaneda} \& {Majda}(1991)]{Majda1991}
{\sc \au{{Avellaneda}, Marco} \& \au{{Majda}, Andrew~J.}} \yr{1991}  \at{{An
  integral representation and bounds on the effective diffusivity in passive
  advection by laminar and turbulent flows}}.  \jt{Communications in
  Mathematical Physics}  \bvol{138}~(2),  \pg{339--391}.

\bibitem[Bhamidipati {\em et~al.\/}(2020)Bhamidipati, Souza \&
  Flierl]{Neeraja2020}
{\sc \au{Bhamidipati, Neeraja}, \au{Souza, Andre~N.} \& \au{Flierl, Glenn~R.}}
  \yr{2020}  \at{Turbulent mixing of a passive scalar in the ocean mixed
  layer}.  \jt{Ocean Modelling}  \bvol{149},  \pg{101615}.

\bibitem[Farrell \& Ioannou(2019)]{Farrell2019}
{\sc \au{Farrell, B.~F.} \& \au{Ioannou, P.~J.}} \yr{2019}  \at{Statistical
  state dynamics: A new perspective on turbulence in shear flow}.  \jt{Zonal
  Jets Phenomenology, Genesis, and Physics. Ed. Boris Galpirin and Peter L.
  Read. Cambridge University Press 2019.}  \pg{pp. 380--400}.

\bibitem[Fernex {\em et~al.\/}(2021)Fernex, Noack \& Semaan]{Fernex2021}
{\sc \au{Fernex, Daniel}, \au{Noack, Bernd~R.} \& \au{Semaan, Richard}}
  \yr{2021}  \at{Cluster-based network modeling\&\#x2014;from snapshots to
  complex dynamical systems}.  \jt{Science Advances}  \bvol{7}~(25),
  \pg{eabf5006},  \arxiv{arXiv:
  https://www.science.org/doi/pdf/10.1126/sciadv.abf5006}.

\bibitem[Flierl \& McGillicuddy(2002)]{FlierlMcGillicuddy}
{\sc \au{Flierl, G.~R.} \& \au{McGillicuddy, D.~J.}} \yr{2002} {\em The Sea:
  Ideas and Observations on Progress in the Study of the Seas\/}, ,
  \vol{vol.~12}, chap. Mesoscale and submesoscale physical-biological
  interactions,  \pg{pp. 113--185}.  \publ{John Wiley and Sons}.

\bibitem[Gallet \& Ferrari(2020)]{Gallet2020}
{\sc \au{Gallet, Basile} \& \au{Ferrari, Raffaele}} \yr{2020}  \at{The vortex
  gas scaling regime of baroclinic turbulence}.  \jt{Proceedings of the
  National Academy of Sciences}  \bvol{117}~(9),  \pg{4491--4497},
  \arxiv{arXiv: https://www.pnas.org/doi/pdf/10.1073/pnas.1916272117}.

\bibitem[Hagan {\em et~al.\/}(1989)Hagan, Doering \& Levermore]{Doering_1989}
{\sc \au{Hagan, Patrick~S.}, \au{Doering, Charles~R.} \& \au{Levermore,
  C.~David}} \yr{1989}  \at{Mean exit times for particles driven by weakly
  colored noise}.  \jt{SIAM Journal on Applied Mathematics}  \bvol{49}~(5),
  \pg{1480--1513},  \arxiv{arXiv: https://doi.org/10.1137/0149090}.

\bibitem[Hassanzadeh {\em et~al.\/}(2014)Hassanzadeh, Chini \&
  Doering]{hassanzadeh2014}
{\sc \au{Hassanzadeh, Pedram}, \au{Chini, Gregory P.} \& \au{Doering,
  Charles R.}} \yr{2014}  \at{Wall to wall optimal transport}.  \jt{Journal of
  Fluid Mechanics}  \bvol{751},  \pg{627–662}.

\bibitem[Hopf(1952)]{Hopf1952}
{\sc \au{Hopf, Eberhard}} \yr{1952}  \at{Statistical hydromechanics and
  functional calculus}.  \jt{Indiana University Mathematics Journal}  \bvol{1},
   \pg{87--123}.

\bibitem[Klus {\em et~al.\/}(2016)Klus, Koltai \& Schütte]{Klus2016}
{\sc \au{Klus, Stefan}, \au{Koltai, Péter} \& \au{Schütte, Christof}}
  \yr{2016}  \at{On the numerical approximation of the perron-frobenius and
  koopman operator}.  \jt{Journal of Computational Dynamics}  \bvol{3},  \pg{51
  -- 79}.

\bibitem[Knobloch(1977)]{knobloch_1977}
{\sc \au{Knobloch, Edgar}} \yr{1977}  \at{The diffusion of scalar and vector
  fields by homogeneous stationary turbulence}.  \jt{Journal of Fluid
  Mechanics}  \bvol{83}~(1),  \pg{129–140}.

\bibitem[Kraichnan(1968)]{Kraichnan1968}
{\sc \au{Kraichnan, Robert~H.}} \yr{1968}  \at{Small‐scale structure of a
  scalar field convected by turbulence}.  \jt{The Physics of Fluids}
  \bvol{11}~(5),  \pg{945--953},  \arxiv{arXiv:
  https://aip.scitation.org/doi/pdf/10.1063/1.1692063}.

\bibitem[Lorenz(1963)]{Lorenz1963}
{\sc \au{Lorenz, E.~N.}} \yr{1963}  \at{Deterministic nonperiodic flow}.
  \jt{Journal of the Atmospheric Sciences}  \bvol{20},  \pg{130--141}.

\bibitem[Maity {\em et~al.\/}(2022)Maity, Koltai \& Schumacher]{Schumacher2022}
{\sc \au{Maity, Priyanka}, \au{Koltai, Péter} \& \au{Schumacher, Jörg}}
  \yr{2022}  \at{Large-scale flow in a cubic rayleigh\&\#x2013;b\&\#xe9;nard
  cell: long-term turbulence statistics and markovianity of macrostate
  transitions}.  \jt{Philosophical Transactions of the Royal Society A:
  Mathematical, Physical and Engineering Sciences}  \bvol{380}~(2225),
  \pg{20210042},  \arxiv{arXiv:
  https://royalsocietypublishing.org/doi/pdf/10.1098/rsta.2021.0042}.

\bibitem[Pope(2011)]{Pope2011}
{\sc \au{Pope, Stephen~B.}} \yr{2011}  \at{Simple models of turbulent flows}.
  \jt{Physics of Fluids}  \bvol{23}~(1),  \pg{011301},  \arxiv{arXiv:
  https://doi.org/10.1063/1.3531744}.

\bibitem[Schneider {\em et~al.\/}(2017)Schneider, Lan, Stuart \&
  Teixeira]{Scheider2017}
{\sc \au{Schneider, Tapio}, \au{Lan, Shiwei}, \au{Stuart, Andrew} \&
  \au{Teixeira, João}} \yr{2017}  \at{Earth system modeling 2.0: A blueprint
  for models that learn from observations and targeted high-resolution
  simulations}.  \jt{Geophysical Research Letters}  \bvol{44}~(24),
  \pg{12,396--12,417},  \arxiv{arXiv:
  https://agupubs.onlinelibrary.wiley.com/doi/pdf/10.1002/2017GL076101}.

\bibitem[Tan {\em et~al.\/}(2018)Tan, Kaul, Pressel, Cohen, Schneider \&
  Teixeira]{Teixeira2018}
{\sc \au{Tan, Zhihong}, \au{Kaul, Colleen~M.}, \au{Pressel, Kyle~G.},
  \au{Cohen, Yair}, \au{Schneider, Tapio} \& \au{Teixeira, João}} \yr{2018}
  \at{An extended eddy-diffusivity mass-flux scheme for unified representation
  of subgrid-scale turbulence and convection}.  \jt{Journal of Advances in
  Modeling Earth Systems}  \bvol{10}~(3),  \pg{770--800},  \arxiv{arXiv:
  https://agupubs.onlinelibrary.wiley.com/doi/pdf/10.1002/2017MS001162}.

\bibitem[Taylor(1922)]{Taylor1920}
{\sc \au{Taylor, G.~I.}} \yr{1922}  \at{Diffusion by continuous movements}.
  \jt{Proceedings of the London Mathematical Society}  \bvol{s2-20}~(1),
  \pg{196--212},  \arxiv{arXiv:
  https://londmathsoc.onlinelibrary.wiley.com/doi/pdf/10.1112/plms/s2-20.1.196}.

\bibitem[Thiffeault(2012)]{Thiffeault2012}
{\sc \au{Thiffeault, Jean-Luc}} \yr{2012}  \at{Using multiscale norms to
  quantify mixing and transport}.  \jt{Nonlinearity}  \bvol{25}~(2),
  \pg{R1--R44}.

\bibitem[Trefethen(2000)]{SpectralMethodsTrefethen}
{\sc \au{Trefethen, Lloyd~N.}} \yr{2000} {\em Spectral Methods in MATLAB\/}.
  \publ{Society for Industrial and Applied Mathematics},  \arxiv{arXiv:
  https://epubs.siam.org/doi/pdf/10.1137/1.9780898719598}.

\bibitem[Venturi {\em et~al.\/}(2013)Venturi, Tartakovsky, Tartakovsky \&
  Karniadakis]{Karniadakis2013}
{\sc \au{Venturi, D.}, \au{Tartakovsky, D.M.}, \au{Tartakovsky, A.M.} \&
  \au{Karniadakis, G.E.}} \yr{2013}  \at{Exact pdf equations and closure
  approximations for advective-reactive transport}.  \jt{Journal of
  Computational Physics}  \bvol{243},  \pg{323--343}.

\bibitem[Weinstock(1969)]{Weinstock1969}
{\sc \au{Weinstock, Jerome}} \yr{1969}  \at{Formulation of a statistical theory
  of strong plasma turbulence}.  \jt{The Physics of Fluids}  \bvol{12}~(5),
  \pg{1045--1058},  \arxiv{arXiv:
  https://aip.scitation.org/doi/pdf/10.1063/1.2163666}.

\end{thebibliography}


\end{document}